# *In-materio* neuromimetic devices: Dynamics, information processing and pattern recognition

Dawid Przyczyna,[1,2,☯] Piotr Zawal,[1,2,☯] Tomasz Mazur,[1,☯] Pier Luigi Gentili[3]\* and Konrad Szaciłowski[1]\*

[1] *Academic Centre for Materials and Nanotechnology, AGH University of Science and Technology, al. Mickiewicza 30, 30-059 Kraków, Poland*

[2] *Faculty of Physics and Applied Computer Science, AGH University of Science and Technology, al. Mickiewicza 30, 30-059 Kraków, Poland*

[3] *Department of Chemistry, Biology, and Biotechnology; University of Perugia, 06123 Perugia, Italy.*

☯ These authors have equally contributed to the manuscript

*E-mail:* pierluigi.gentili@unipg.it, szacilow@agh.edu.pl

**Abstract**

The story of information processing is a story of great success. Todays' microprocessors are devices of unprecedented complexity and MOSFET transistors are considered as the most widely produced artifact in the history of mankind. The current miniaturization of electronic circuits is pushed almost to the physical limit and begins to suffer from various parasitic effects. These facts stimulate intense research on neuromimetic devices. This feature article is devoted to various *in materio* implementation of neuromimetic processes, including neuronal dynamics, synaptic plasticity, and higher-level signal and information processing, along with more sophisticated implementations, including signal processing, speech recognition and data security. Due to vast number of papers in the field, only a subjective selection of topics is presented in this review.

# 1. Introduction

The story of information processing is a story of great success. Todays' microprocessors are devices of unprecedented complexity and MOSFET transistors are considered as the most widely produced component in the history of mankind, with a total number of devices approaching 13 sextillions ($1.3 \times 10^{22}$) of fabricated devices.[1] The contemporary microprocessor contains approximately $3.95 \times 10^9$ transistors.[2] In a human, there are an estimated 10–20 billion neurons in the cerebral cortex and 55–70 billion neurons in the cerebellum.[3] A human brain is, therefore, the most complex information processing structure, as each of the $6 \times 10^{10}$ neurons may form up to $10^4$ synaptic connections with other neurons. Brain's ability to learn and adapt is a consequence if its dynamically changing topology of synaptic connections, plasticity of individual connections, high redundancy and multilevel dynamics at various geometrical and temporal scales.

These facts stimulate intense research on neuromimetic devices. Their performance, at the present stage of development, cannot be compared with natural systems, but also provide stimulation for other fields of investigation, including chemistry, physics, electronics, and computer sciences.

This feature article is devoted to various *in materio* implementation of neuromimetic processes, including neuronal dynamics, synaptic plasticity, and higher-level signal and information processing. From the plethora of various *in-materio* implementation of information processing, involving inorganic and organic materials, polymers, various molecular species, as well as biopolymers and even living organisms[4] we have chosen a handful of wet photochromic systems and semiconducting materials. This selection is by no means exhaustive, but sufficient to illustrate the main research directions as well as current trends in *in-materio* neuromimetic computing.

# 2. Mimicking neural dynamics

Human intelligence emerges from the complex structural and dynamical properties of our nervous system. The primary cellular elements of our nervous systems are neurons. The ultimate computational power of our nervous system relies on the dynamical properties of neurons and their networks. Every neuron is a nonlinear dynamic system,[5] and according to some theoretical analysis can be regarded as a biological memristive element.[6-8] Some neurons operate in the oscillatory regime. They are called pacemaker neurons and fire action potentials periodically. Pacemaker neurons generate rhythmic activities in neural networks involved in the neocortex, basal ganglia, thalamus, locus coeruleus, hypothalamus, ventral tegmentum area, hippocampus, and amygdala.[9] These structures are associated with sleep, wakefulness, arousal, motivation, addiction, memory consolidation, cognition, and fear.

Excitable neurons are another type of neurons present in the nervous system. Excitability can be twofold – "tonic" or "phasic". When neurons react to a constant excitatory signal by firing a sequence of spikes, they are classified as "tonic". On the other hand, excitability is "phasic", when neurons react in an analog manner and shoot only once, when receiving a sharp excitatory signal. Excitatory "tonic" neurons are present e.g. in the cortex whereas "phasic" excitable neurons act e.g. in the auditory brainstem (involved in precise timing computations) and in the spinal cord.[10, 11]

Finally, there are chaotic neurons. Chaotic neurons are quite common in the nervous system because the intrinsic dynamic instability facilitates the extraordinary ability of neural networks to adapt.[12, 13]

It is possible to mimic the dynamics of neurons by selecting specific chemical systems and maintaining them out-of-equilibrium.[14] One of the most widespread examples is the Belousov-Zhabotinsky reaction. Other popular models are memristive elements and circuits,[6-8] as well as some (photo)electrochemical systems, especially those with self-excitable oscillations.[15]

## *2.1. The case of the Belousov-Zhabotinsky reaction*

The Belousov-Zhabotinsky (BZ) reaction is a catalyzed oxidative bromination of malonic acid in aqueous acidic solution (1):

$2BrO_3^-{}_{(aq)} + 3CH_2(COOH)_2{}_{(aq)} + 2H^+{}_{(aq)} \rightarrow 2BrCH(COOH)_2{}_{(aq)} + 3CO_2{}_{(g)} + 4H_2O_{(l)}$ (1)

Various metal ions or metal-complexes, such as either cerium ions or ferroin (i.e. tris-(1,10-phenanthroline)-iron(II)) or tris(2,2'-bipyridyl)dichloro-ruthenium(II) (Ru(bpy)$_3^{2+}$) can serve as catalyst. The mechanism of the BZ is quite complicated because it consists of many elementary steps. Briefly, when the concentration of the intermediate bromide ($Br^-$) is higher than its critical value, the reaction proceeds by a set of elementary steps wherein the catalyst maintains the reduced state, and the solution is red-colored in the presence of ferroin. During these elementary steps, bromide is consumed. As soon as the concentration of $Br^-$ is lower than its critical value (that corresponds to 5x10$^{-6}$ [$BrO_3^-$]), the reaction proceeds through another set of elementary steps, where mono-electronic transformations are involved, and the catalyst goes from the reduced to the oxidized state. In the presence of the indicator ferroin, the solution becomes blue. When the concentration of the oxidized state of the catalyst becomes high, another set of reactions becomes important where bromide is produced. As soon as the bromide concentration becomes again higher than its critical value, the solution switches from blue to red, and the cycle repeats. A BZ reaction in its reduced state is like a resting neuron in its hyperpolarized state. When the BZ reaction feels a small perturbation, it maintains its reduced state. On the other hand, when the perturbation is sufficiently strong, slightly above a critical threshold value, it responds by moving temporarily to its oxidized state and then recovering its original reduced state. By a careful choice of the contour conditions (i.e., concentrations of the reagents, temperature, and flow rate, in case the reaction is performed in an open system), it is possible to have the BZ reaction in either the oscillatory or the tonic excitable or the chaotic regime.[16] The BZ reaction in the oscillatory regime is a good model of real pacemaker cells. Pacemaker cells have their internal rhythm, but external stimuli can alter their timing. In pacemaker cells, information about a stimulus is encoded by changes in the timing of individual action potentials, and it is used to rule proprioception and motor coordination for running, swimming and flying.[17] In a similar way to neurons, the BZ reaction in the oscillatory regime can be perturbed in its timing by both inhibitors and activators. Bromide is an example of an inhibitor, whereas Ag$^+$ is an example of an activator, or more precisely of an anti-inhibitor because its addition removes bromide, forming an AgBr precipitate. The effect of injection of either $Br^-$ or Ag$^+$ is immediate, and the BZ reaction restores the initial period quickly, after one or a few more cycles, if it is carried out in an open system such as a Continuous-flow Stirred Tank Reactor (CSTR). The CSTR guarantees a replenishment of fresh reagents and the elimination of the products. The response of the system is phase-dependent, where for the phase of addition we mean the ratio

$\varphi = \frac{\tau}{T_0}$ (2)

In (2), τ is the "time delay", i.e., the time since the most recent spike occurred, and $T_0$ is the period of the previous oscillations. The addition of $Br^-$ leads always to a delay in the appearance of a spike. In other words, $\Delta T = T_{pert} - T_0$ ($T_{pert}$ is the period of the perturbed oscillation) is always positive. The higher the phase of the addition of $Br^-$, the larger the ΔT.[18] The addition of silver ion decreases the period unless it is injected in small quantities and at a low phase, inducing a slight lengthening of the period of oscillations.

Since the information within our brain is encoded as a pattern of activity of neural networks, it is compelling to study the coupling between artificial neurons and the corresponding dynamics. The coupling among real neurons takes place through discrete chemical pulses of neurotransmitters released by the synapses of a neuron and collected by the dendrites of other neurons. Therefore, it is useful to focus on pulse-coupled oscillators. The study of two pulse-coupled BZ oscillators, implemented in two physically separated CSTRs, may be viewed as the chemical analog of the two pulse-coupled pacemaker cells. The dynamics of two pseudo-neurons has been investigated under both symmetrical inhibitory and/or excitatory coupling.[19, 20] The latency caused by the propagation of action potentials has been emulated employing a delay between the appearance of a spike and the release of a pseudo-neurotransmitter. Both symmetrical and asymmetrical coupling can give rise to many temporal patterns. For example, mutual and symmetrical inhibitory coupling generates either anti-phase, in-phase or irregular oscillations depending on the time delay and concentration of the inhibitor; when $\tau$ is zero and $[Br^-]$ is large, suppression of oscillations in one artificial neuron model – that is maintained in its reduced state – has been observed. Even the mutual and symmetrical excitatory coupling generates different dynamical regimes: the so-called master and slave condition, bursting behavior, fast anti-phase oscillations, and suppression of oscillations with the suppressed oscillator maintained in its oxidized state. Further patterns have been achieved with mixed excitatory-inhibitory coupling and with the symmetrical coupling of two unequal BZ oscillators.[21] All these dynamical patterns emulate the reasoning code of pairs of real neurons. However, the main drawback of these artificial systems is their hybrid nature. The chemical coupling is ruled by a silicon-based computer (see Figure 1). To contrive chemical oscillators that can couple autonomously, it is useful to focus on optical signals and photo-sensitive oscillators.

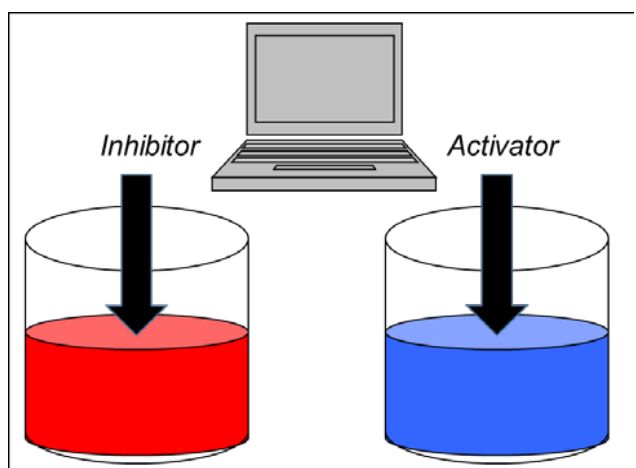

*Figure 1*. Scheme of a computer-controlled coupling between two BZ reactions performed in two distinct CSTRs.

### *2.2. Optical Communication among Artificial Neuron Models.*

The oscillatory BZ reaction with cerium ions as catalysts gives rise to appreciable transmittance oscillations in the UV part of the electromagnetic spectrum (see Figure 2a). If a UV radiation with constant intensity crosses the BZ reaction, its transmitted intensity is modulated. In other words, the BZ reaction transmits a UV output, whose intensity oscillates, and the frequency of oscillations coincides with the intrinsic frequency of the BZ. When ferroin is introduced as either the catalyst or the redox indicator, large transmittance oscillations are recordable in all the visible region of the spectrum (see Figure 2B). Therefore, the BZ reaction becomes suitable to transmit a visible oscillatory signal. Transmittance oscillations in the visible spectrum and oscillatory red luminescence are recordable in case $Ru(bpy)_3^{2+}$ is chosen as the catalyst (see Figure 2C).

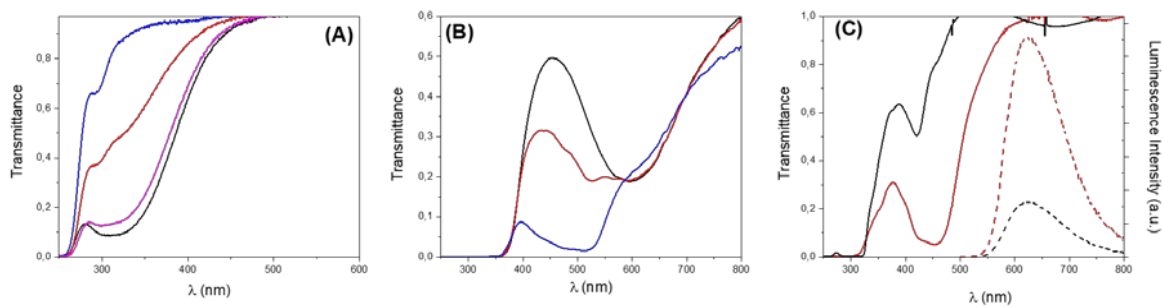

*Figure 2. Transmittance oscillations for the BZ with cerium ions (a), with ferroin (b), and with*

*Ru(bpy)$_3$$^{2+}$ (c). The dashed spectra in (c) represent the oscillations in the red luminescence of [Ru(bpy)$_3$]$^{2+}$.*

The periodic UV or visible radiation, transmitted or emitted by the BZ reaction, can be sent to luminescent and photochromic compounds.[22] Luminescent and thermally reversible photochromic compounds are good models of phasic excitable neurons. They respond to a steady excitatory signal with analog output of emitted light or color saturation. Moreover, they "relax", i.e., they recover the initial state upon cessation of the excitatory signal. When the BZ reaction transmits a periodic excitatory optical signal to the photo-excitable receiver, constituted by either a luminescent or a photochromic compound, a master-and-slave relationship is always established. The light emitted by the luminescent compound or the saturation of the color generated by the photochromic compound oscillate with the same frequency of the BZ. The oscillations of the optical signals of the transmitter and the receiver are in-phase or out-of-phase depending on the response rate of the "slave": if it is fast, the synchronization is in phase, whereas if it is slow, the sync is out-of-phase.

Photochromic materials find further applications in neuromimetic systems based on optical signals. Photo-reversible photochromic compounds allow to implement memory effects: if they are direct photo-reversible photochromes, UV and visible signals promote and inhibit their colorations, respectively. Furthermore, when a photochromic compound receives excitatory optical signals at the bottom of a liquid column, wherein there are either laminar or turbulent convective motions of the solvent generated by a vertical thermal gradient, it gives rise to a Hydrodynamic Photochemical Oscillator that originates chaotic spectrofluorimetric signals.[23-25] A Hydrodynamic Photochemical Oscillator can mimic a chaotic neuron. If it sends its chaotic excitatory signal to a luminescent compound, the latter synchronizes in-phase and emits an aperiodic fluorescence signal having the same chaotic features of the transmitter.[26]

Finally, the intrinsic spectral evolution of every photochromic compound that transforms from one form to the other under irradiation, generates either positive or negative feedback actions. The optical feedback actions produced by every photochromic compound act on both itself and other photo-sensitive neuro-mimetic systems that are optically connected to the photochrome. Therefore, recurrent networks can be implemented by using photochromic compounds.[22] The feedback actions of every photochrome are wavelength-dependent because its photo-excitability, which depends on the product $\varepsilon\Phi$, where $\varepsilon$ is its absorption coefficient and $\Phi$ is its photochemical quantum yield, is also wavelength-dependent. Therefore, photochromic compounds allow implementing neuromodulation,[27] which is the alteration of neuronal and synaptic properties in the context of neuronal circuits, allowing anatomically defined circuits to produce multiple outputs reconfiguring networks into different functional circuits.[28]

When photochromic and luminescent compounds are combined with luminescent oscillatory reactions, such as the chemiluminescent Orbán transformation, they allow implementing even feed-forward networks wherein optical signals travel unidirectionally.[22] The recurrent and feed-forward networks implemented so far with oscillatory and photo-excitable chemical systems consist of two or, at most, three nodes that communicate through an optical code. The UV and visible signals can play both excitatory and inhibitory effects. Within the networks, phenomena of in-phase, out-of-phase, anti-phase, and phase-locking synchronizations have been observed as it occurs in networks of real neurons that communicate through the chemical code of neurotransmitters. Three communication architectures have been devised, labeled as α, β, and γ, respectively. In the α architecture, the transmitter and the receiver are in the same cuvette and the same phase, possibly with one component chemically protected by micelles. In the β architecture, the transmitter and the receiver are in the same cuvette but two immiscible phases. In the γ architecture, the transmitter and the receiver are in two distinct cuvettes. The networks have been attained by hybridizing or upgrading the α, β, and γ architectures.[22] All these optical communications do not require an external source of information, such as computer software, to be guided, but they are spontaneous and maintained



## 2.3. An artificial neuron in photoelectrochemical system

Various photoelectrochemical processes have been considered as a computational platform for quite a long time.[29] The basis of these processes was a photoelectrochemical photocurrent switching effect (PEPS)[30, 31] observed in numerous surface-modified semiconductors[32] as well as in highly defected bulk semiconductors like cadmium sulfide,[33] bismuth sulfoiodide[34] and bismuth oxyiodide.[35] The effect was utilized for the implementation of various binary logic gates, reconfigurable logic gates,[36, 37] combinatorial logic circuits[38-41] and arithmetic systems.[42] Later the PEPS-based devices were used to implement ternary logic functions,[43] ternary combinatorial circuits[44] and fuzzy logic systems.[44, 45] The analysis of the PEPS effect and its applications were based solely on the thermodynamics of modified semiconducting materials, whereas the kinetic aspects of photocurrent generation were neglected. The involvement of kinetic factors leads to new switching phenomena with potentials applications in information processing.[46]

Detailed photoelectrochemical studies of cadmium sulfide – multiwalled carbon nanotubes composite have shown a peculiar dynamic behavior of this material subjected to short pulses of light.[47] The photoelectrodes prepared from this composite material, when subjected to a series of light pulses, generate photocurrent pulses, the intensity of which depends on the history of the photoelectrode. In more detail, it depends on past illumination history: the second pulse yields a photocurrent of higher intensity when the interval between pulses is short enough. The response of the electrode towards trains of pulses of various frequencies are shown in Figure 3. Quite surprisingly, the dependence of the photocurrent spike intensity on time interval between pulses is described by a bi-exponential equation, like in the case of living plastic neurons.[48]

Hebbian learning[49] is based on the plastic properties of synapses. Synaptic plasticity is the process that strengthens or weakens the connection between neurons as a consequence of the time sequence of firing events. It was figuratively described as "neurons wire together if they fire together".[50] This process is usually described by bi-exponential equations (3):[51, 52]

$$W(\Delta t) = \begin{cases} A_+ \exp(-\Delta t / \tau_a) + |A_-| \exp(-\Delta t / \tau_c) & \text{for} \quad \Delta t \geq 0 \\ A_+ \exp(\Delta t / \tau_b) + |A_-| \exp(\Delta t / \tau_d) & \text{for} \quad \Delta t < 0 \end{cases}, \quad (3)$$

where $\Delta t$ is the time interval between post- and presynaptic signals, $\tau_\pm$ are the time constants and $A_\pm$ are the parameters determined by the synaptic weights. The upper part of the formula (the $\Delta t \geq 0$ case) described the potentiation mode (i.e. an increase of synaptic weight as a consequence of the decreased time interval between events), whereas the lower one (the $\Delta t < 0$ case) the depression mode. In the described case only the potentiation mode has been observed, therefore the response function can be simplified to (4):[47]

$$f(\Delta t) = \alpha_1 \exp(-\Delta t / T_1) + \alpha_2 \exp(-\Delta t / T_2) + \beta \quad (4)$$

where $\alpha_{1,2}$ and $T_{1,2}$ are relevant fitting parameters while $\beta$ is only a scale factor, which is irrelevant for the data interpretation. The result of the fitting procedure is presented in Figure 3c. A good match was found ($\chi^2$ around $1.1 \times 10^{-3}$) with the following set of fitting parameters: $\alpha_1=3.014\pm0.034$, $\alpha_2=4.80\pm0.21$, $T_1=116.4\pm3.7$ ms, $T_2=6.88\pm0.31$ ms, and $\beta=1.722\pm0.024$. It is noteworthy that the time constants are of comparable value to the ones obtained for biological structures.[51, 53]



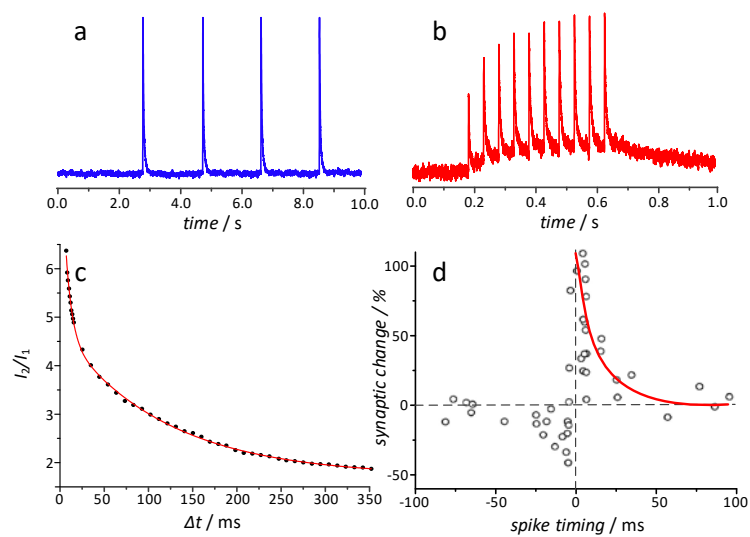

***Figure 3.*** *The response of the artificial synapse upon illumination (450 nm) with 2 s (a) and with 50 ms (b) time intervals between light pulses. The plasticity of the studied synaptic system with the fit line described by equation (2) (b) is compared to the plasticity of hippocampal glutamatergic neurons in culture (d) (data are taken from Ref.* [48]*). Reproduced from Ref.* [47] *with permission from Wiley.*

The mechanism of the plastic processes observed in the CdS-MWCNT composite is relatively simple. Photoexcitation of the material within the fundamental absorption of cadmium results in the promotion of electrons from the valence to the conduction band (process **1** in Figure 4). Some electrons may undergo thermal relaxation to the valence band (**1'**). The observed photocurrent is a consequence of interfacial electron transfer to the conducting substrate (**2**) accompanied by the redox reaction with the redox mediator (iodide anions in the studied case) present in the electrolyte (**2'**). In the presence of carbon nanotubes, acting as additional electron traps the process (**3**) competes with photocurrent generation (**2**). The recombination process (**3'**) must be significantly slow, therefore the lifetime of trapped electrons is in order of tenths of a second. Partial filling of the nanotube-related trap states results in increased photocurrent intensities.

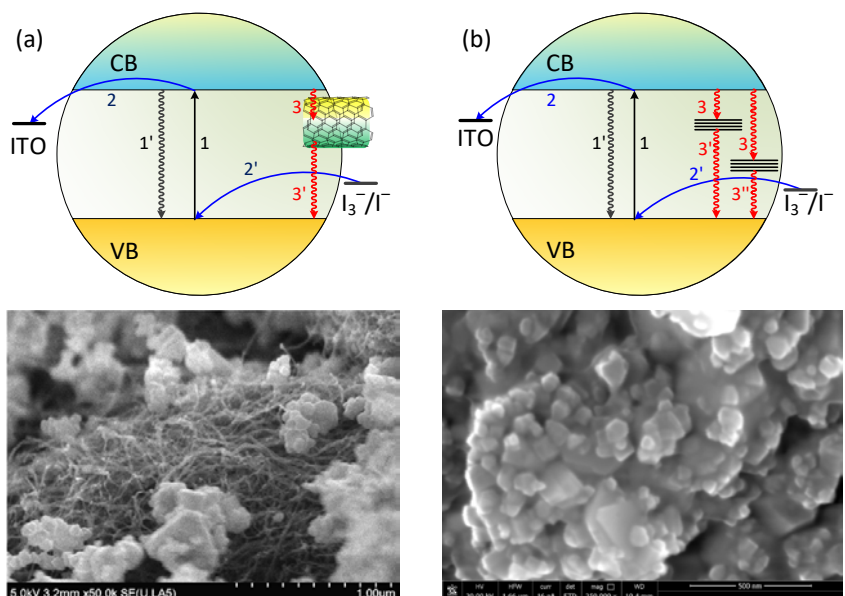

***Figure 4.*** *The mechanism responsible for the synaptic behavior of the CdS/MWCNT-based device (a) and hexagonal/tetragonal CdS mixtures (b) along with SEM images of studied materials. Adapted from Refs.* [47, 54]*.*



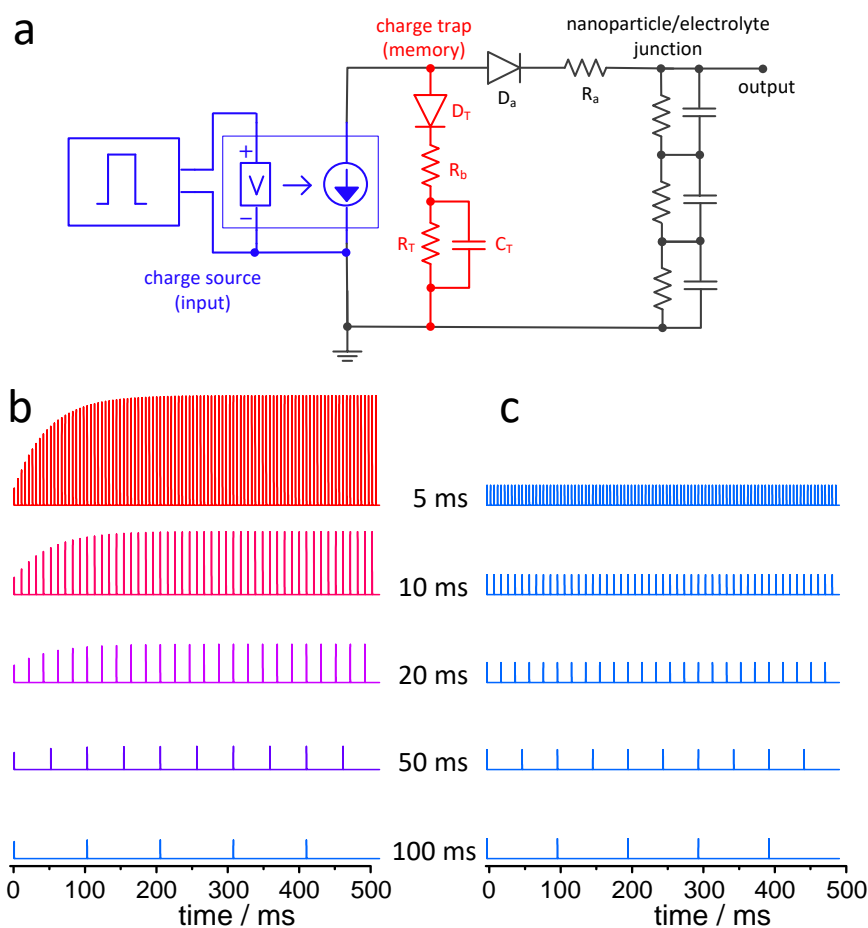

*Figure 5.* The equivalent electronic circuit employed to describe the ITO/MWCNTs/CdS photoelectrodes (a) and the simulated time-dependent response of the equivalent circuit to the pulsed stimulation characterized by various repetition intervals. Two cases were taken into consideration: in the first one the branch responsible for charge trapping/de-trapping events (related to the presence of MWCNTs) was introduced (b) and in the second one it was neglected (c). Reproduced from Ref. [47] with permission from Wiley.

The SPICE model (Figure 5a) proves the correctness of the proposed mechanism. Photoelectrode based on the CdS-MWCNT material was modeled by three RC loops, which simulate the behavior of a nanoparticulate electrode as suggested in previous reports.[31, 55] The nanotube trapping site was modeled by an additional $R_T C_T$ loop. The capacitance of this part was set to be 50 times larger (500 nF) than the ones used in the other RC loops. This choice is justified by a very high electric capacitance of carbon nanotubes.[56] The equivalent circuit contains also other elements: $R_a$ describes the ohmic resistance of the electrolyte, whereas $R_b$ is the electron transfer resistance of the CdS-MWCNT junction. Two diodes in the circuit represent Schottky junctions between conducting support and the material ($D_a$) and between carbon nanotubes and semiconductor nanoparticles ($D_T$). These diodes provide unidirectional electron transfer from the conduction band to the conducting substrate ($D_a$) and from the conduction band to the nanotube traps ($D_T$).

The application of change pulses results in trains of output potential of characteristics very close to the experimental ones (cf. Figure 5b). Removal of the charge trapping subcircuit (red in Figure 5) results in a disappearance of the learning effect (Figure 5c). Very recently it was found that similar effects can be observed in other highly defected nanocrystalline CdS modifications. Very similar photoelectrochemical neuromimetic devices were based on a 1:1 mixture of hexagonal and tetragonal polymorphs of cadmium sulfide (Figure 4b). In this particular case, the bi-exponential



learning curve was also observed ($\chi2 = 6.23 \times 10^{-5}$), but the fitting parameters were significantly different ($\alpha_1$=0.218±0.013, $\alpha_2$=0.339±0.014, $T_1$=167±23 ms, $T_2$=20±4 ms, and $\beta$=1.00±0.01).[54]

## 3. Processing Fuzzy Logic with molecules

Human intelligence has the remarkable power of handling both accurate and vague information. Vague information is coded through the words of our natural languages. We have the remarkable capability to reason, speak, discuss, and make rational decisions without any quantitative measurement and any numerical computation, in an environment of uncertainty, partiality, and relativity of truth. A major challenge is the design of neuromimetic devices that have the capabilities of human intelligence to compute with words.[57] The imitation of the human ability to compute with words is still challenging. One of the approaches that can offer a satisfying approximation is Fuzzy logic-based models.

### 3.1. Some key concepts of Fuzzy logic

Fuzzy logic has been defined as a rigorous logic of vague and approximate reasoning.[58] Fuzzy logic is based on the theory of fuzzy sets proposed by the engineer Lotfi Zadeh in 1965.[59] A Fuzzy set is different from a classical Boolean set because it breaks the law of Excluded Middle. An item may belong to a Fuzzy set and its complement at the same time, with the same or different degrees of membership. The degree of membership (µ) of an element to a Fuzzy set can be any number included between 0 and 1. It derives that Fuzzy logic is an infinite-valued logic. Fuzzy logic can be used to describe any non-linear cause and effect relation by building a Fuzzy Logic System (FLS). The construction of an FLS requires three fundamental steps. First, the granulation of all the variables in Fuzzy sets. The number, position, and shape of the Fuzzy sets are context-dependent. Second, the graduation of all the variables: each Fuzzy set is labeled by a linguistic variable, often an adjective. Third, the relations between input and output variables are described through syllogistic statements of the type "If…, Then….", which are named as Fuzzy rules.

The "If…" part is called the antecedent and involves the labels chosen for the input fuzzy sets. The "Then…" part is called the consequent and involves the labels chosen for the output fuzzy sets. When we have multiple inputs, these are connected through the AND, OR, NOT operators.[60] In formulating the fuzzy rules, we must consider all the possible scenarios, i.e., all the possible combinations of input fuzzy sets. At the end of the three-steps procedure, an FLS is built; it is a predictive tool or a decision support system for the particular phenomenon it describes.

Every FLS is constituted by three components: A Fuzzifier, a Fuzzy Inference Engine, and a Defuzzifier. A Fuzzifier is based on the partition of all the input variables in fuzzy sets. It transforms the crisp values of the input variables in degrees of membership to the input fuzzy sets. The Fuzzy Inference Engine is based on fuzzy rules. It turns on all the rules that involve the fuzzy sets activated by the crisp input values. Finally, the Defuzzifier is based on the fuzzy sets of the output variables, and it transforms the collection of the output fuzzy sets, activated by the rules, in crisp output values. Fuzzy logic is a good model of the human ability to compute with words because there are some structural and functional analogies between any FLS and the Human Nervous System.[61]

### 3.2. The Fuzziness of the Human Nervous System

The Human Nervous System (HNS) comprises three elements: (I) the sensory system; (II) the central nervous system; (III) the effectors' system. The sensory system catches physical and chemical signals and transduces them in electro-chemical information that is sent to the brain. In the brain, information is integrated, stored, and processed. The outputs of the cerebral computations are electro-chemical commands sent to the components of the effectors' system, i.e., glands and muscles.[62] The sensory system includes visual, auditory, somatosensory, olfactory, and gustatory subsystems. Each type of sensory subsystem encodes four features of a stimulus: modality, intensity, time evolution, and spatial distribution.



The power of distinguishing these features derives from the hierarchical structure of every sensory system. In fact, for each sensory system, we have, at the smallest level, a collection of distinct molecular switches. At an upper level, we have a set of distinct sensory cells: each cell contains many replicas of a specific molecular switch. Finally, at the highest level, we have many replicas of the different receptor cells that are organized in a tissue whose structure depends on the architecture of the sensory organ. For instance, in the case of the visual system, we have four types of photoreceptor proteins (each one absorbing a specific portion of the visible spectrum); four types of cells (one rod and three types of cones), each one having many replicas of one of the four types of photoreceptor proteins. Finally, we have many replicas of the different cells disposed on a photo-sensitive tissue, the retina, with the rods spread on the periphery, and the cones concentrated in the fovea.

Consequently, the information of a stimulus is encoded hierarchically. The collection of four types of photoreceptor proteins plays like an ensemble of four distinct molecular fuzzy sets. The information regarding the modality of the stimulus is encoded as degrees of membership of the stimulus to the four molecular fuzzy sets; that is, it is encoded as fuzzy information at the molecular level ($\bar{\mu}_{ML}$). The four types of cells play like cellular fuzzy sets. The information regarding the intensity is encoded as degrees of membership of the stimulus to the cellular fuzzy sets, that is, as fuzzy information at the cellular level ($\bar{\mu}_{CL}$). Finally, the array of the many replicas of the receptive cells plays like an array of cellular fuzzy sets, and the information regarding the spatial distribution of the stimulus is encoded as degrees of membership to the array of cells, that is, as fuzzy information at the organ level ($\bar{\mu}_{OL}$). The total information of the stimulus will be the composition of the fuzzy information encoded at the three levels. For instance, in the case of the photoreceptor system, the total information will be a matrix of data reproducing the array of cells on the retina. Each element of the matrix will be the product of two terms: the fuzzy information encoded at the molecular level times that encoded at the cellular level.

The sensory cells produce graded potentials that are analog signals. The information of such signals is usually converted into the firing rate of the action potential trains. Often, the action potentials are produced by an architecture of afferent neurons that integrate the information regarding the spatial distribution of the stimuli. Every afferent neuron has a receptive field that works as a fuzzy set encompassing specific receptor cells.[63] The action potentials generated by the afferent neurons are the ideal code for sending the information up to the brain. In the cerebral cortex, various areas are having different intrinsic rhythms.[64] They form a neural dynamic space partitioned in overlapped cortical fuzzy compartments. Such cortical fuzzy sets are activated at different degrees by separate attributes of the perceptions and produce a meaningful experience of the external and internal worlds.

### *3.3. The best strategies to implement Fuzzy Logic Systems (FLSs)*

In electronics, the best implementations of FLSs have been achieved through analog circuits, although fuzzy logic is routinely processed in digital electronic circuits. More recently, fuzzy logic has also been processed by using molecules, macromolecules, and chemical transformations. All the methods proposed for processing fuzzy logic can be sorted out in three main strategies.[63] The first strategy is an imitation of the "fuzzy parallelism" of the sensory subsystems described in the previous paragraph. The second is the "conformational fuzziness" of molecules and macromolecules that exist as an ensemble of conformers, whose distribution is context-dependent. The third is the "quantum fuzziness" that hinges on the decoherence of overlapped quantum states originating continuous, smooth, analog input-output relationships between macroscopic variables when it involves large amounts of molecules. In the next paragraphs, examples of the three strategies are described.

### *3.3.1. The "Fuzzy Parallelism" of the Biologically Inspired Photochromic Fuzzy Logic Systems.*

As we have seen in paragraph 3.2, the human visual system grounds on four photoreceptor



proteins: three for daily vision in color and one for night vision in black and white. All of them have 11-*cis* retinal as the chromophore. However, the four photoreceptors have different absorption spectra in the visible region, because they differ in the amino-acidic composition. The absorption spectra of the four photoreceptor proteins behave as "molecular fuzzy sets". The spectral composition of a light stimulus is encoded as degrees of membership of the light to these "molecular fuzzy sets". Moreover, the millions of replicas of the three photoreceptor proteins within each photoreceptor cell allow determining the intensity of the signals at every wavelength. The imitation of the way we distinguish colors has allowed devising chemical systems that extend human vision to the UV.[65] Such chemical systems are based on direct thermally reversible photochromic compounds. Direct photochromic species usually absorb just in the UV. The criteria to mix the direct photochromic compounds and generate Biologically Inspired Photochromic Fuzzy Logic (BIPFUL) systems that extend the human ability to distinguish electromagnetic frequencies to the UV region have been the following ones. First, the absorption bands of the closed uncolored (*Un*) forms were assumed to be input fuzzy sets. Second, the absorption bands of the open colored (*Col*) forms were assumed to be output fuzzy sets. Third, the algorithm expressing the degree of membership of the UV radiation, having the intensity $I_0(\lambda_{irr})$ at the wavelength $\lambda_{irr}$, to the absorption band of the *i*-th compound is given by (5):

$$\mu_{UV,i} = \Phi_{PC,i}(\lambda_{irr}) I_0(\lambda_{irr}) \left(1 - 10^{-\varepsilon_{Un,i} C_{0,i} l}\right) \quad (5)$$

In equation (5), $\Phi_{PC,i}(\lambda_{irr})$ is the photochemical quantum yield of photo-coloration, $\varepsilon_{Un,i}$ the absorption coefficient at $\lambda_{irr}$ for the *i*-th photochromic species, and $C_{0,i}$ is its analytical concentration. Finally, the equation expressing the activation of the *i*-th output fuzzy sets is (6):

$$A_{Co,i} = \frac{\varepsilon_{Co,i}(\lambda_{an})}{k_{\Delta,i}} \mu_{UV,i} \quad (6)$$

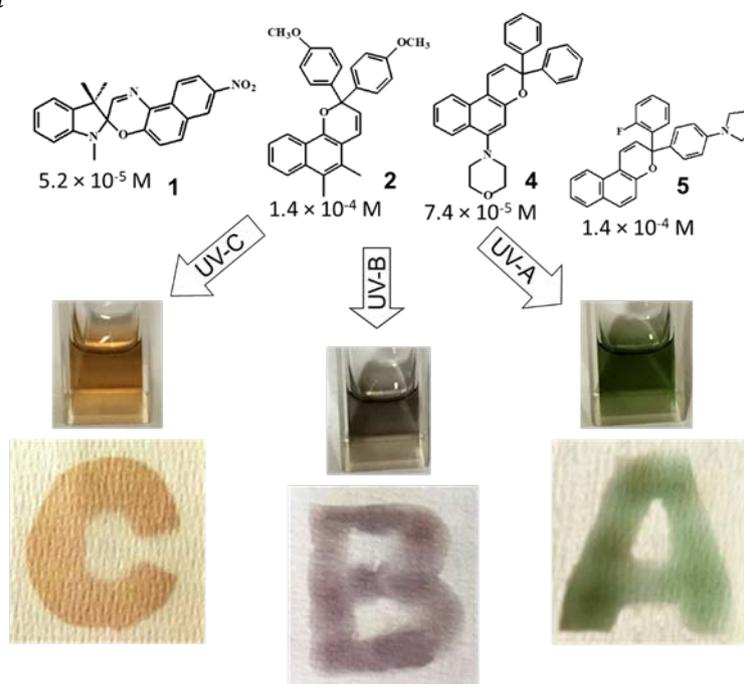

***Figure 6***. *A quaternary BIPFL system constituted by the direct thermally reversible photochromic compounds **1**, **2**, **4**, and **5**; this BIPFUL system becomes green, grey, and orange when it is irradiated by frequencies belonging to the UV-A, UV-B, and UV-C regions, respectively. Its sensory activity works both in the liquid phase and on a white paper made of cellulose.*

In equation (6), $A_{Co,i}$ is the absorbance at the wavelength $\lambda_{an}$ into the visible and due to the colored form of the *i*-th photochromic species; $\varepsilon_{Co,i}(\lambda_{an})$ is its absorption coefficient, and $k_{\Delta,i}$



is its kinetic constant of the bleaching reaction. Each absorption spectrum recorded at the photo-stationary state will be the sum of as many terms represented by equation (6) as there are photochromic components present within the BIPFUL system. Some BIPFUL systems consisting of from three to five photochromic compounds have been proposed.[66] They allow the three regions of the UV spectrum, UV-A, UV-B, and UV-C, to be discriminated because the wavelengths belonging to the three UV regions originate from distinct colors. An example of a quaternary BIPFUL system is shown in Fig 6. It is constituted by four direct thermally reversible photochromic compounds (labeled as **1**, **2**, **3**, and **5**).[65] It becomes green, gray, and orange when it is irradiated by frequencies belonging to the UV-A, UV-B, and UV-C regions, respectively. It works both in the liquid phase (the concentrations of the species involved are reported in Figure 6) and on a cellulosic white paper.

The imitation of other sensory subsystems, described as hierarchical fuzzy systems, wherein distinct molecular and cellular fuzzy sets work in parallel, will allow the design of new artificial sensory systems. These new artificial sensory systems will have the power to extract the essential features of stimuli and will contribute to the recognition of variable patterns.

*3.3.2. The Fuzziness of conformers*

Every molecular or macromolecular compound that exists as an ensemble of conformers works as a fuzzy set.[63, 67] The types and amounts of different conformers depend on the physical and chemical contexts. Every compound is like a word of the natural language, whose meaning is context-dependent. Conformational dynamism and heterogeneity enable context-specific functions to emerge in response to changing environmental conditions and allow the same compound to be used in multiple settings. It is possible to quantify the fuzziness of every compound by determining the fuzzy entropy (7):

$$H = -\frac{1}{\log(n)} \sum_{i=1}^{n} \mu_i \log(\mu_i) \qquad (7)$$

wherein $n$ is the number of conformers and $\mu_i$ is the relative weight of the *i*-th conformer.

The fuzziness of a macromolecule is usually more pronounced than that of a simpler molecule. Among proteins, those completely or partially disordered are the fuzziest.[68] Their remarkable fuzziness makes them multifunctional and suitable to moonlight, i.e., play distinct roles, depending on their context.[69]

*3.3.3 From quantum to fuzzy logic*

The elementary unit of quantum information is the qubit. The qubit, $|\Psi\rangle$, is a quantum system that has two accessible states, labeled as $|0\rangle$ and $|1\rangle$, and it exists as a superposition of them (8):

$$|\Psi\rangle = a|0\rangle + b|1\rangle \qquad (8)$$

In equation (8), $a$ and $b$ are complex numbers that verify the normalization condition $|a|^2 + |b|^2 = 1$. The two states, $|0\rangle$ and $|1\rangle$, work as two fuzzy sets. The $|\Psi\rangle$ state belongs to both $|0\rangle$ and $|1\rangle$ with degrees that are $|a|^2$ and $|b|^2$, respectively. Any logic operation on a qubit manipulates both states, simultaneously. If a molecular system is a superposition of n qubits, any operation on it manipulates $2^n$ states, simultaneously. Therefore, it is evident the alluring parallelism of quantum logic. However, deleterious interactions between the quantum system and the surrounding environment can cause the decoherence of the quantum states.[70] The decoherence induces the collapse of any qubit in one of its two accessible states, either $|0\rangle$ or $|1\rangle$, with probabilities $|a|^2$ and $|b|^2$, respectively.

Whenever the decoherence is unavoidable, the single microscopic units can be used to process discrete logics, i.e., binary or multi-valued crisp logics depending on the original number of qubits.[71, 72] Advanced microscopic techniques, reaching the atomic resolution, are required to carry out the computations with single particles. Alternatively, large assemblies of particles, e.g., molecules, can be used to make computations. However, vast collections of molecules (amounting to the order of the



Avogadro's number) are bulky materials. The inputs and outputs for making computations become macroscopic variables that can change continuously. The functions linking input and output variables can be either steep or smooth. Steep sigmoid functions are suitable to implement discrete logic. In contrast, both linear and nonlinear smooth functions are suitable to build fuzzy logic systems.[73]

Many fuzzy logic systems have been built by using the emission of light as preferable output because it bridges the gap between the microscopic and the macroscopic world. For instance, the fluorescence of 6(5H)-phenanthridinone depends smoothly on the hydrogen bonding donation ability of the solvent (HBD) and the temperature.[74] The fluorescence of tryptophan, both as an isolated molecule and bonded to the serum albumin, depends smoothly on the temperature and the amount of the quencher flindersine.[75] Further examples are a ruthenium complex, whose fluorescence depends on $Fe^{2+}$ and $F^-$,[76] and europium bound to a metal-organic framework, which depends on metal cations, such as $Hg^{2+}$ and $Ag^+$.[77]

With a multi-responsive chromogenic compound, belonging to the class of spirooxazine, all the fundamental fuzzy logic gates, AND, OR, and NOT, have been implemented.[78] The protons, $Cu^{2+}$, and $Al^{3+}$ ions have been used as inputs and the color coordinates (R, G, B) or the colorability of the chromogenic compound as outputs. Then, other platforms have been proposed. For example, a multi-state tantalum oxide memristive device[79] and an anthraquinone-modified titanium dioxide electrode.[44] All these case studies demonstrate that fuzzy logic can be processed by unconventional chemical systems showing analog physical-chemical input-output relationships in either the liquid or the solid phase. They are alternative to the conventional way of processing fuzzy logic, which is based on electronic circuits and signals.

## 4. Classification and transformation of simple signals

All materials based real devices operating at realistic conditions ae best described by fractional differential equations, as it was demonstrated in the case of a capacitor by Svante Westerlund in 1991 in a seminal paper with a mind-twisting title "Dead matter has memory!".[80] This concept was further extended towards other fundamental devices.[81, 82] It implies, that *in-materio* components exhibit some forms of memory, which is a consequence of internal dynamics. Therefore these systems and devices are naturally suited for signal processing and also can be incorporated, as active nodes, into signal classification devices. The following sections will present some selected applications in this field, which also relate to neuromorphic information processing.

### *4.1. Generation of higher harmonics in memristive devices*

The second harmonic generation (SHG) involves generating signals (e.g. optical or electrical) the frequencies of which are twice as high as the fundamental frequency, hence often this effect is called frequency doubling. The generation of higher harmonics is observed in non-linear resistors, but in this case, the largest spectral weight falls on the fundamental frequency.[83] Electronic circuits capable of implementing electrical frequency doubling is a diode bridge. It can be shown based on Fourier analysis that the diode bridge achieves 4.5% efficiency for SHG and 18.9% for higher harmonics in relation to input power.[84] Oskoee et al.[85] initially suggested the potential for SHG for strongly memristive systems. In an attempt at improving SHG efficiencies, Cohen et al.[83] performed a quantitative analysis based on the memristor model for a single element as well as for the memristor bridge. The results show a significant improvement in performance in SHG generation, at 16.9% for a single element and 40.3% for a memristor bridge. Based on the above simulation results, the potential of memristive structures in applications related to SHG has been shown.

Literature describing research on SHG present in hardware memristive structures is sparse. Majzoub et. al. performed an analysis of the influence of higher harmonic components of the recorded current on the pinched hysteresis loop.[86] Authors used a commercially available device (KNOWM Inc.) to show, that higher harmonic components are crucial to form the pinched hysteresis loop. By filtering the components above second harmonic, authors obtained distortionless response of the



device, without losing functionality in digital applications of AND/OR logic gates simulation. To benefit from the higher complexity of the system and get as close as possible to high interconnectivity of biological nervous structure Avizienis et al.[87] have studied the neuromorphic atomic switch network (ASN). Presented ASN was fabricated through electroless self-assembly of silver nanowires from the $AgNO_3$ solution which was added to SU-8 (epoxy-based negative resist) with patterned Cu posts. The oxidation reaction of Cu seeds with the $AgNO_3$ solution leads to the formation of an extremely interconnected structure ($10^9$ junctions per $cm^2$) of silver nanowires with variable morphology. Gas-phase sulfurization of the obtained network enables the formation of memristive metal-insulator-metal ($Ag|Ag_2S|Ag$) junctions. Analysis of the network properties showed the memristive nature of the junctions after activation through unidirectional electrical sweeps which were associated with the formation of conducting filaments. Moreover, studied ASN exhibited a pronounced increase in the magnitude of higher harmonic generation (Figure 7) after system functionalization due to an increase in the number of hard switching junctions.[85]

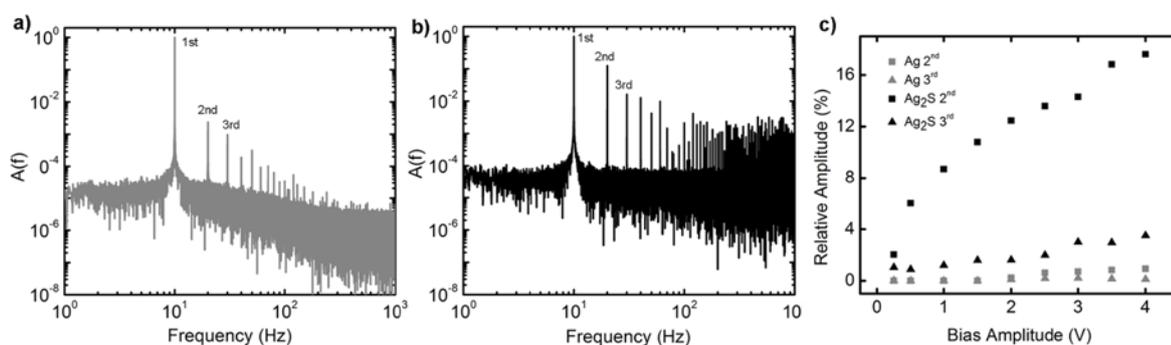

*Figure 7. Fourier transform spectra for control device (a) and functionalized network (b), both subjected to 10Hz, 2V stimulation. Second and third harmonic relative amplitudes in the function of bias voltage show a significant increase in SHG for the functionalized network (black) in comparison with the control device (grey). Reproduced from Ref. [87] with permission from the Public Library of Science.*

### *4.2. Amplitude discrimination*

The amplitude discrimination of voltage is the inherent property of memristors. Only pulses with sufficient amplitude can affect the resistive state of the memristive device. In general, the resistive state is dependent on the history applied current and voltage, enabling storage and adjustable changes of the device's conductance. Specific voltage pulse can transfer the memristor from one state (e.g. high resistance state) to another (e.g. low resistance state).

Memristors employed in reservoir system can act as a simple amplitude classifier. In general, the resistive state of the memristive device is dependent on the history of the applied current and voltage. Furthermore, exploiting the non-linear I-V dependence, one can gradually switch the resistance to the other state with the pulses of sufficient amplitude and proper polarity. These inherent properties of memristors can be utilized for the amplitude classification with employing the memristor in the feedback loop. It has been reported that lead(II) iodide ($PbI_2$) incorporated in a single echo-state network with delayed feedback loop can efficiently discriminate input voltage pulses based on their amplitude.[88] The $PbI_2|Cu$ device shows a distinct rectifying characteristic, which leads to amplification of forward bias pulses and reduction of reverse bias pulses. In the reservoir system, the output signal – after a delay time – is routed back as input, which evaluates to one iteration. As the number of feedback cycles rose, some signals increased and the others were attenuated, leading to an amplitude classification. If the signal voltage exceeded the threshold value of 1.85 $V_{pp}$, the signal was amplified. On the contrary, lower voltage amplitudes were decreased over iterations. As another example, $[SnI_4((C_6H_5)_2SO)_2]/Cu$ in echo state machine can perform similar classification based on both amplitude and duration of the input voltage.[89] Higher the amplitude and longer the duration, the signal propagated longer in the feedback loop before the full attenuation.



## *4.3. Frequency discrimination*

Most basic frequency discrimination for neuromorphic memristor-based devices is spike-rate dependent plasticity (SRDP) learning strategy. If the frequency of spike train exceeds the threshold value, the short term effects, such as signal intensity decay can be overcome and each next spike has increased intensity, until reaching plateau level. This way of frequency processing has its instant limits – a range of frequencies between threshold and plateau values. The resolution of frequency detection is also associated with individual pulse length. The realization of these ideas is possible with the incorporation of several memristive materials, for example, organolead trihalide perovskites (or organic-inorganic perovskites).[90] Even transistors, such as nanoparticles/organic memory transistors which are equivalent to leaky memory devices and have kind of STP-like characteristics can be used as very simple frequency discriminators.[91, 92] Presented strategy is unfortunately insufficient for a wide variety of frequencies and lacks scientific elegance – one still needs dedicated software and von-Neumann architecture-based hardware elements to detect measure and correctly categorize output signals.

Advancement of the above methodology would be to use STP or LTP (short- and long-time plasticity) effects associated with threshold frequency. Experimentally this behavior was shown by He in 2014[93] for the sandwich-like structure of Pt/FeO$_x$/Pt. To observe this effect one must put the spike train similar to the biological firing curve. The spikes are inspired by the firing behavior of biological neurons – and consist of two pulses. The first one is a very short but high-amplitude pulse, followed by a wide and low-amplitude pulse in the opposite direction. It is a necessary condition to realize bidirectional weight change. At the same time, this is the limitation of the method as the input impulses must be properly shaped in order to distinguish between their frequency. Low-frequency spikes (below 10 kHz) tent to induce long-term depression (LTD) by decreasing the conductivity of the memristor. On the contrary, 20 kHz spikes accumulate positive constituting pulses, causing an increase of the conductance – cf. Figure 8 a and b.

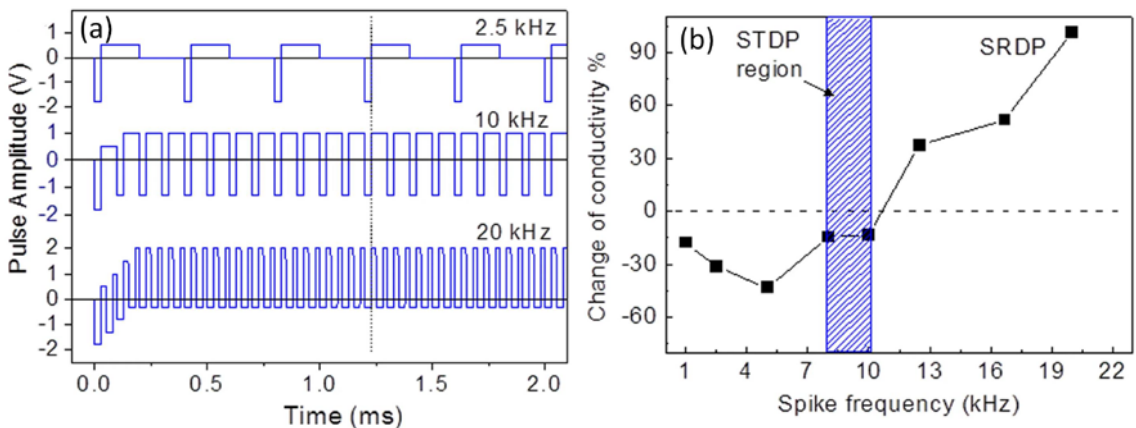

*Figure 8*. *a) Spike waveforms, similar to biological neural spikes at different presynaptic firing frequency. Signal intensification is detected after crossing the frequency threshold. b) Emulated SRDP learning rule for memristor build of iron oxide. Such behavior is also reported for real-life synapses.*

Similarly, better frequency resolution than mere SRDP can also be obtained by directly using the learning rules of Bienenstock–Cooper–Munro (BCM) according to whom the synapse's weight can exhibit either be strengthened (potentiation) or weakened (depression) even when subjected to the same spike trains. Memristive devices with WO$_x$ between metal (Pt) electrodes additionally to STP/LTP regimes utilize the so-called "sliding threshold frequency" as the frequency classification rule.[94] Historical synaptic activity influences device performance – periods of increased activity are followed by higher frequency threshold for synaptic weight potentiation and periods of lighter intensity   –as presented in Figure 9a for biological neurons.



This behavior was repeated artificially (Figure 9b) – after 200 Hz stimulation, synaptic weight increased and the concomitant 10 Hz stimulation caused current drop (step 2 in Figure 9b). After that 1 Hz stimulation and 10 Hz stimulation caused increase in output current (step 4 in Figure 9b). Figure 9c shows results for stimulation with different frequency (20, 50 and 100 kHz) and then five probing pulses. After stimulation these probing pulses, depending in turn on their own frequency, result in either weakening or strengthening of the synaptic weight. The higher the pre-stimulation is before probing pulses, the higher frequency probing pulse is needed to strengthen the synapse – effect is similar to the one biologically observed. In addition to frequency discrimination, pretreatment of the sample influences also amplitude threshold (Figure 9d).

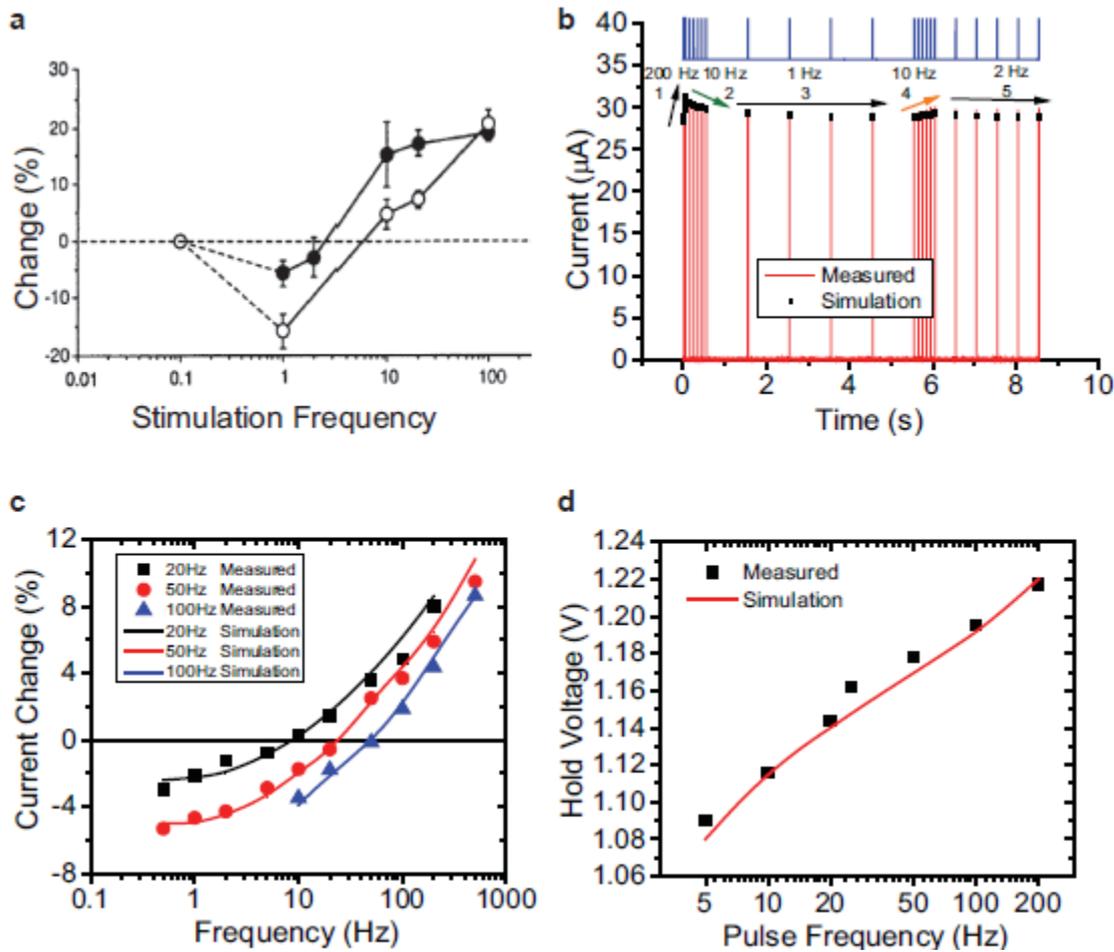

*Figure 9. Activity-dependent plasticity (STP/LTP rules) and sliding threshold effects in memristors. Relative change in synaptic weight as a function of stimulation frequency for two different cases. Low stimulation frequency results in depression and high stimulation frequency results in facilitation. The threshold moves to a lower frequency for filled symbols (low activity period) compared to the normal condition (open symbols). Data obtained in rat visual cortex [95] (a). Memristor response to consecutive programming pulse trains at different frequencies. The 10 Hz pulse train caused a decrease of current in step 2 following strong stimulation in step 1, but the current increase in step 4 following weak stimulation in step 3. Black squares: Simulation results from the memristor model using experimental parameters (b). Memristor current change as a function of the stimulation frequency after the memristor has been exposed to different levels of activities. Pulse trains consisting of five pulses (1.2 V, 1 ms) with different repetition frequencies were used to program the memristor. Black squares, red circles, and blue triangles represent experimental data. The solid lines are simulation results from the memristor model using experimental parameters (c) Measured threshold hold voltage as a function of previous activity*



*(represented by different pulse frequency). The device was stimulated by pulse trains with the same repetition frequency of 50 Hz but different amplitudes. Black squares: Experimental data. Solid line: Simulation results from the memristor model using experimental parameters (d). Reprinted from Ref. [94] with the permission of Wiley.*

The learning/forgetting effect with the frequency threshold was also reported for ZnO memristive devices.[96] Described devices possessed a biorealistic rate-dependent synaptic plasticity, mimicking biological systems, has been demonstrated in the rectifying diode-like Pt/n-ZnO/SiO$_{2-x}$/Pt synaptic heterostructures. Among others, SRDP rule, STP and LTP retention, and frequency sliding threshold simultaneously exist in the device. The PPF phenomenon along with the SRDP learning rule was discovered to similarly follow the plasticity behavior of that in the actual synapse. The frequency sliding threshold was explored to show the dynamic stability of the synaptic weight depending on spike train time spacing and frequency. In whole, frequency discrimination signal processing help emulated human-like "Learning-Forgetting-Relearning" synaptic behavior. These findings will serve as cornerstones for dynamic hardware-based neuromorphic systems.

Discrimination of periodic signals of various waveforms according to their frequency should be also possible. It is an obvious fact that due to the switching dynamics memristors are frequency-sensitive elements. At sufficiently high frequencies they may behave like linear memristors, whereas their nonlinear features emerge as low-frequency range. Therefore, it should be possible to build a reservoir system similar to the previously described, which selectively amplifies signals of frequencies lower than the threshold value and attenuates signal of higher frequencies (or vice versa). Such behavior is not a unique feature in classical analog electronics: any bandpass filter (active or passive) can perform similarly. The expected advantage of reservoir memristive devices in comparison to analog filters is a very high slope of their frequency characteristics, especially if they are embedded in a single node echo state machine[89, 97] as shown in Figure 10.

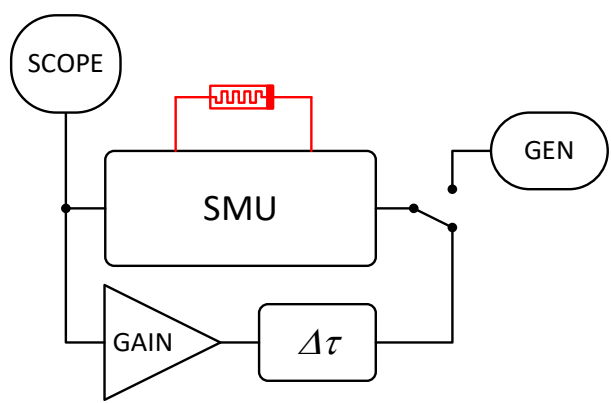

*Figure 10. A scheme of a memristor-based single node echo machine: SMU stands for a source-measure unit, GAIN for an amplifier, SCOPE for oscilloscope/signal recorder Δτ for a delay line and GEN for arbitrary function generator.*

In such systems, low-frequency signals should be amplified, whereas the signals of frequencies higher than the characteristic cut-off frequency should be slowly attenuated. The signal matching the cut-off frequency should remain unchanged. The same type of devices can be also used for more advanced signal processing due to gradual change in the signals' symmetry due to partial rectification at the Schottky junction of the memristor.

Some of the above-described effects were implemented for memristive devices and perform frequency-discrimination functions, amplitude-discrimination functions alongside with other time-oriented functions in patented solutions.[98] The sum of output currents, thus overall resistivity of the aggregated memristors network, is dependable on the input signal frequency level – making it



possible to switch the device between states if the input signal is greater than a threshold frequency.

## 5. Classification of complex acoustic patterns

Dynamics of resistance changes in memristors as well as their highly nonlinear characteristics seem to be key features in their application is signal processing and classification. Furthermore, they can be incorporated into feedback loops yielding single-node echo state machines (or other types of reservoir computers) with a superb performance in signal classification. In the case of reservoir computing training of the reservoir is not required, the classification of the input signal relies on the internal dynamics of the reservoir. The only point that requires training is a readout layer -–simple artificial neural network (software-based), single layer perceptron or a simple signal processing circuit. Despite obvious utility of memristive elements in such computational tasks the reports on experimental verification of memristor applicability in signal classification/processing are scarce, however, the number of theoretical works, including numerical simulations, is increasing. The reason is purely technological – analog memristors (vide infra) are an emerging class of devices and require a lot of fundamental and technological studies.

There are two principal categories of memristors, which can be tentatively called analog and digital ones.[99, 100] Analog memristors gradually change their internal state due to interfacial switching processes (e.g. Schottky barrier height modulation) or dopant migration. As a result, these devices can store not only binary data but also analog data. Materials that show the interface-switching behavior are still under development. Moreover, the accuracy in controlling the memristance value in analog memristors is still considered to be a big concern, however, recent studies on pulse and signal classification define the safe limits of their applicability.[89, 97, 101] On the contrary, most memristors are known that they are based on the filamentary-switching mechanism. In filamentary switching, memristors can have either a high resistance state (HRS) or a low resistance state (LRS) and therefore the systems based on these devices are considered error-tolerant. On the other hand, the application of binary memristors significantly limits the computational performance of memristive systems. The computational power of memristive systems can be further improved by utilization of their dynamic properties, e.g. in reservoir computer systems.[102]

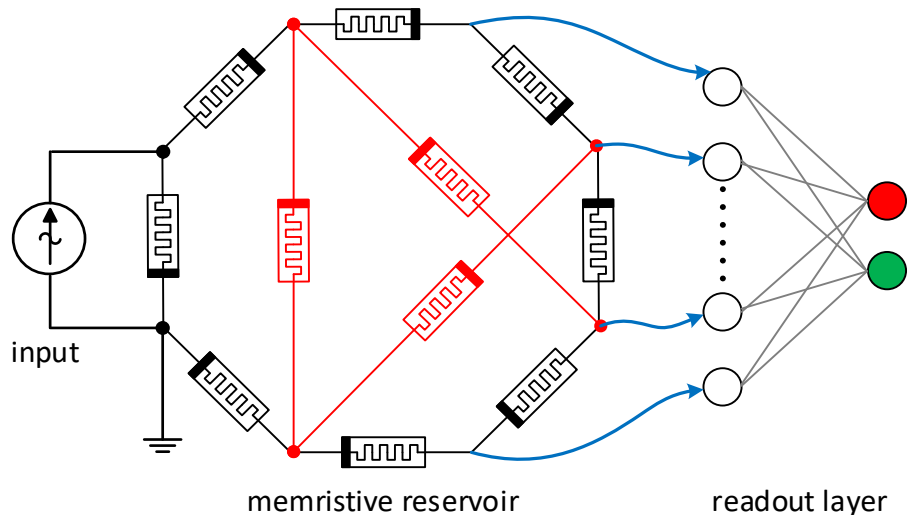

*Figure 11. Schematic representation of a memristive reservoir computing system in two different topologies: ring (only black memristors) and small-world (black and red memristors). In the studied case the readout layer was a one later perceptron with a sigmoidal activation function. Adapted from Ref. [103].*

The concept of signal classification based on the dynamic behavior of memristor was reported by Tanaka et al.[103] In this report authors demonstrate the applicability of linear-drift-based memristors in discrimination between sine and triangular waves of the same amplitude within a



refined range of frequencies. The system used two different topologies of reservoir networks: ring and small-world topology (a network, in which most nodes are not neighbors of one another, but the neighbors of any given node are likely to be neighbors of each other and most nodes can be reached from every other node by a small number of hops or steps), as shown in Figure 11.

Such a system computes an output potential at each node and these values are fed into two nodes of a perceptron. The input of a perceptron node $h_i$ ($i = 1, 2$) is a weighted sum of inputs from reservoir node (9):

$$h_i(t) = \sum_j w_{i,j} x_j(t) \qquad (9)$$

Sigmoidal activation function was used to compute the output state of the perceptron nodes (10):

$$y_i(t) = \frac{1}{1+e^{-h_i(t)}} \qquad (10)$$

Training of the reservoir involves optimization of weights to achieve the (0,1) output state for sine wave input and the (1,0) for the triangular one. The systems provided good separability of the waveforms at a sufficiently high number of reservoir nodes coupled with the perceptron: 2 in the case of ring topology and 5 in the case of small-world topology. On the other hand, the ring topology works efficiently only in the case of identical elements, even a small variation in memristor characteristics significantly reduces the performance of the system. The introduction of additional connectivities in the circuit (small-world topology) results in a variability-tolerant system, however, a larger number of readout connections (blue arrows in Figure 11) are required for optimal performance.

Similar capabilities, even without the trained readout layer should be also observed in a single node echo state machine with a non-linear node of appropriate characteristics. Composite waveforms, with Fourier spectra covering a significantly large range of frequencies, should yield a complex dynamic behavior: some spectral components should be amplified, whereas some others attenuated. This may lead to a binary classification of waveform shapes. It can be further extended by an appropriate readout layer.

### *5.1. Classification of musical objects*

Music is the most ubiquitous human activity independently on any social and cultural attributes or intellectual abilities, however, according to some opinions, it does not convey any biologically-relevant information.[104] According to Guerino Mazzola music provides a platform of communication between symbolic and emotional layers.[105] Music, like information, is a notion very difficult to define in precise terms. Dislike speech, music is not meant for explicit communication purposes, but it triggers various emotional responses in recipients due to aesthetical feelings. On the other hand, music is a very well organized structure, as not every combination of sounds should be considered as music, however, the modern musicological approach provides a piece of evidence that any purposeful combination of sounds can be considered as music.[106, 107] Going to the extreme, even silence (a lack of purposeful sound) can be considered music, with famous *4'33"* by John Cage as the most prominent example.[108] In the simplest approach, however, music can be defined as an appropriate time sequence of quantized frequencies (Figure 12). These frequencies are called steps in a musical scale, and along with rhythm and timbre are principal constituents to any musical piece. In Western music, an octave (an interval between frequencies f and 2f) is divided into 12 steps, called semitones, but other musical systems use quartertones (the Middle East and India) or other smaller intervals. Musical harmony is a complex notion reflected in: (i) the pure content of the ensemble of frequencies heard at given time (also including a timbre of an individual note), (ii) the musical content - the verticality if the chord (a set of notes played simultaneously) and (iii) the position and relation of a chord in relation to the melody at given moment.[109, 110] Despite well-established musical theories[109] the automated classification of intervals, chords and clusters and recognition of



consonance and dissonance has been not achieved. Furthermore, the understanding of physical nature of dissonance and consonance is still not fully established,[104] however various approaches beyond classical Helmholtz curve has been developed.[111-113]

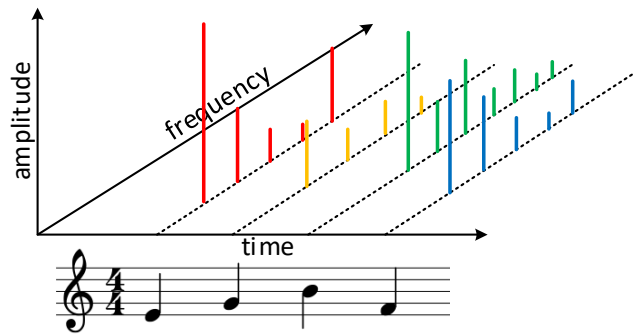

*Figure 12. 3D representation of music as a time sequence of tones of different frequency (pitch) and spectral characteristics (timbre). Adapted from Ref. [109].*

Photoelectrochemical reservoir systems, based on wide bandgap semiconductors (neat or modified with simple coordination compounds) cannot compete with photonic devices in terms of speed or efficiency. They operate, however, in a frequency domain corresponding to the audible range. Therefore, we have turned our attention to the classification of acoustic signals. There are a plethora of different data sets that require advanced processing techniques, including ECG and EEG signals, automated speech analysis or the classification of music. We have found the latter as the most suitable one to be addressed in the photoelectrochemical system due to its internal, well-defined structure based on the Pythagorean geometry.[104, 109, 114]

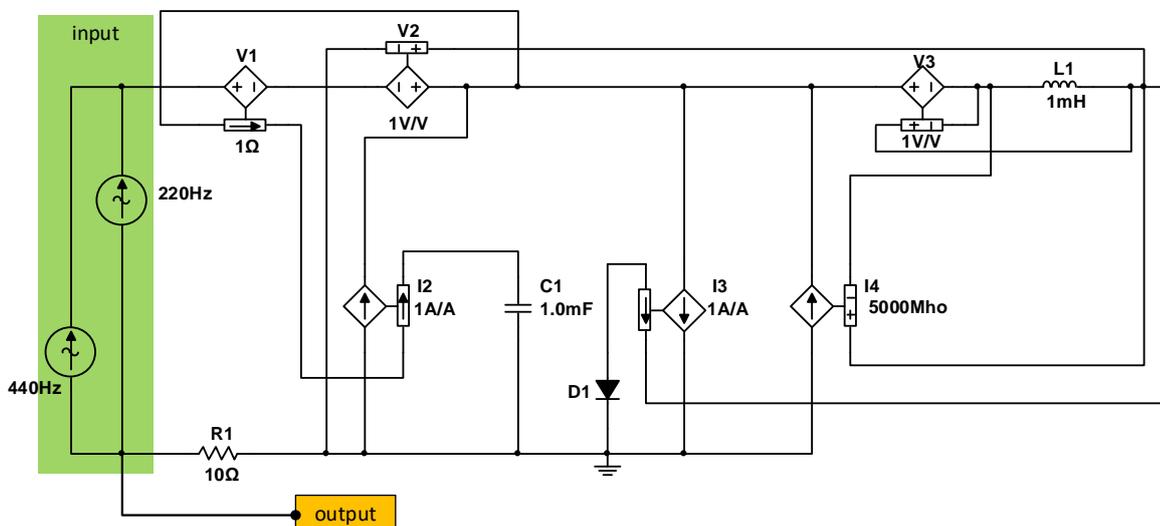

*Figure 13. Circuit by Song et al. transforming a diode into a memristor.[115] Two sine-wave current sources are used as an input, whereas the voltage drop on resistor R1 is used as an output.*

Before the experimental verification of this idea series of numerical simulations for a simple memristive device (a memristor in series with a resistor) have been performed. As the numerical model, an example given by Song et al. was implemented in Multisim (Figure 13), in which a nonlinear circuit element (a diode) is converted into memristor by mutator circuit based on a capacitor and an inductor connected by a series of voltage- and current-controlled voltage- and current sources.[115] It can be observed, that pure tones (single sine waves of frequencies which belong to the natural scale) yield simple memristor-like pinched hysteresis loops (Figure 14). When a sum of



two signals, which form a consonant combination (e.g. perfect octave, perfect fifth or perfect fourth) a stable loop is observed as well, but with an increased number of lobes (Figure 15). In the case of an interval which is considered dissonant (e.g. triton), the observed characteristics becomes quasi-random (the simulation was too short to justify if the signal is truly chaotic or not), but the trajectory is confined inside the hysteresis loop defined by the lower frequency tone (Figure 15 c). Therefore, the memristor hysteresis loop can be considered as an attractor for the unstable behavior of the memristor-based circuit.

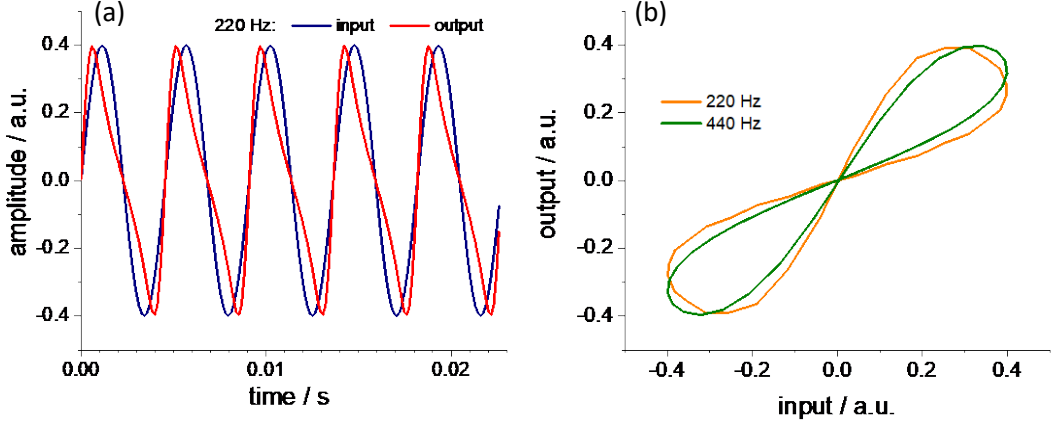

*Figure 14.* Simulated responses of a resistor-memristor circuit subjected to the voltage signal modulated with 220 Hz sine (a) and corresponding I-V (input/output) hysteresis loop for 200 and 440 Hz sine waves (b).

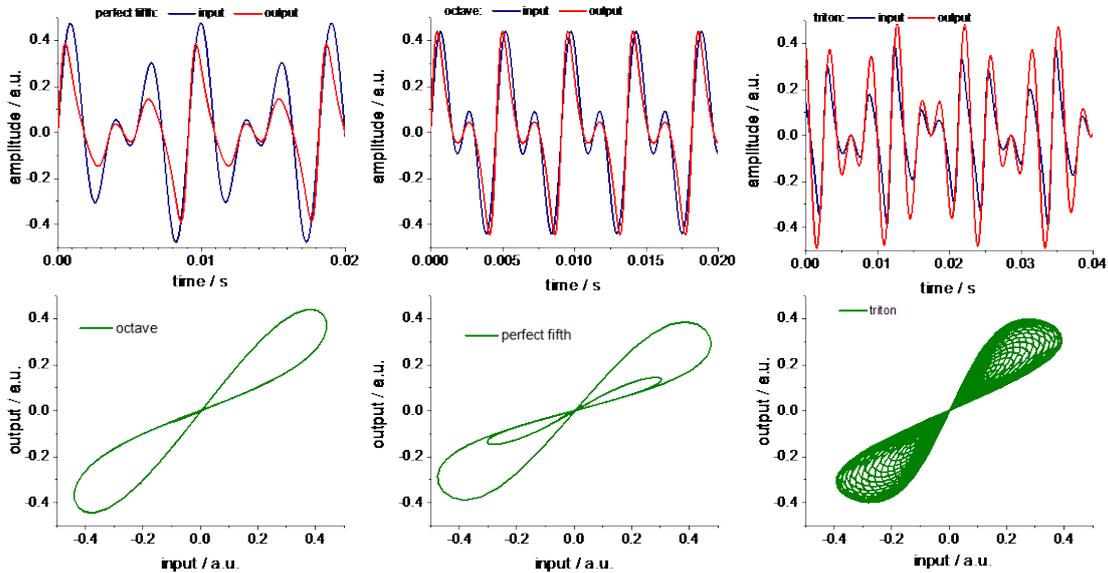

*Figure 15.* Simulated responses of a resistor-memristor circuit subjected to the voltage signal modulated with two sine waves: 220 Hz + 440 Hz (octave, a), 220 Hz + 330 Hz (perfect fifth, b) and 220 Hz + 309.375 Hz (triton, c).

This result is fully consistent with the chaotic behavior of memristive circuits, especially the so-called Chua circuit (a double loop circuit of a memristor, resistor, and two capacitors).[116] This numerical experiment proves the utility of reservoir computing in the processing of acoustic signals, which has been recently postulated on the basis of theoretical models.[117] Surprisingly, the results are consistent with the results of neurophysiological studies on the perception of music by humans and monkeys.[118, 119] In conclusion, this experiment opens a new path into a field of interdisciplinary



investigations: the application of molecular systems to the analysis, in the short term, but maybe also to the creation of music in the future. This idea has been recently successfully developed by Professor Eduardo R. Miranda in a series of biocomputing experiments with *Physarum* slime mold[120-123] and follows the cross-boundary research at the interface of the information theory, music, and physical sciences.[124] The analogy between the music is perceived by humans and the reservoir perception of simple intervals may be misleading. It does not mean that a simple reservoir is as sensitive as a human ear, but rather it may suggest that memristive systems may provide a universal problem-solving power and with appropriate operation can solve numerous problems, which cannot be easily addressed using other approaches. The combination of reservoirs with logic devices (Boolean or fuzzy) may lead to a substantial increase of complexity and computing efficiency. The first reports on practical combinations of the Boolean logic and the reservoir computing are already available.[125, 126]

### *5.2. Speech processing and classification*

Speech recognition is a fundamental yet complex problem for AI systems of the modern era. Speech is the most common means of communication among the human race, therefore the automated speech recognition system finds numerous applications. Verbal communication is a trivial task in everyday life, but this becomes a complex phenomenon when ported to the machines. The complexity of this task originates from an extremely reach vocabulary of a single language (hundreds of thousands of words), variations of the pronunciations and dialects of the same words, and variations in a timbre, rhythm of speech and personal characteristics. The speech of children and non-native speakers adds additional complications to this already very complicated task.[127] Therefore, automated speech recognition is a complex problem in the field of artificial intelligence.[128] Software solutions include Fourier and wavelet analysis followed by artificial neural network-based classification of spectral features.[129] The most promising approach, however, is based on neuro-inspired speech recognition, involving reservoir computing.[130-132] Although there are many advances reported on this front with software simulations, the solutions are not scalable to port it to the hardware of an intelligent machine. Therefore, hardware-based solutions based on memristors and other nonlinear elements are considered as potential candidates to embed acoustic frequency signal analysis and classification (vide supra). An addition of oscillatory characteristics (or other dynamic features) should increase the performance of the computing system based on small networks.[133, 134]

Up to now, there is only one reported experimental evidence of successful human speech recognition in a hardware system. The device reported by Romera et al.[135] is based on four spin-torque oscillators based on the $PtMn/Co_{71}Fe_{29}/Ru/Co_{60}Fe_{20}B_{20}/Co_{70}Fe_{30}/MgO/Fe_{80}B_{20}/MgO/Ta/Ru$ magnetic tunnel junctions fabricated by ultrahigh vacuum magnetron sputtering. The FeB layer presents a structure with a single magnetic vortex as the ground state. In a small region called the vortex core (of about 12 nm diameter at remanence for our materials), the magnetization spirals out of plane. Under direct current injection and the action of the spin transfer torques, the core of the vortex steadily gyrates around the center of the dot with a frequency in the range of 150 MHz to 450 MHz for the oscillators reported in the cited paper.[135] A set of English vowels was used as an input. Because usually the formant frequencies of a human voice are within the range of 500-3500 Hz, and the magnetic nano-oscillators have the characteristic frequencies in the MHz range, the formant frequencies were used to synthesize the pair of input signals ($f_A$, $f_B$) as linear combinations of three principal formant frequencies. These high-frequency signal $I_{RFA}$ and $I_{RFA}$, calculated on the basis of 37 female voices, served as input into neuromimetic classification device (Figure 16 a, b). The device itself contains four oscillating magnetic tunneling junctions connected into a circuit driven with DC currents to maintain radio frequency oscillations. Microwave signals recorded as an output and used for vowel



recognition (Figure 16 d, e). The correlation maps involving two input signal were generated (Figure 16 f) for various vowels represented on $f_A$, $f_B$ plane (Figure 16 g).

The performance of this system was compared with a multilayer perceptron trained to use 12 formant frequencies. The software-based speech recognition tool, with ca. 100 trained weights yields a recognition rate of 97%. In comparison, the oscillatory network with four oscillators yields a recognition rate of 84% with 30 training parameters.

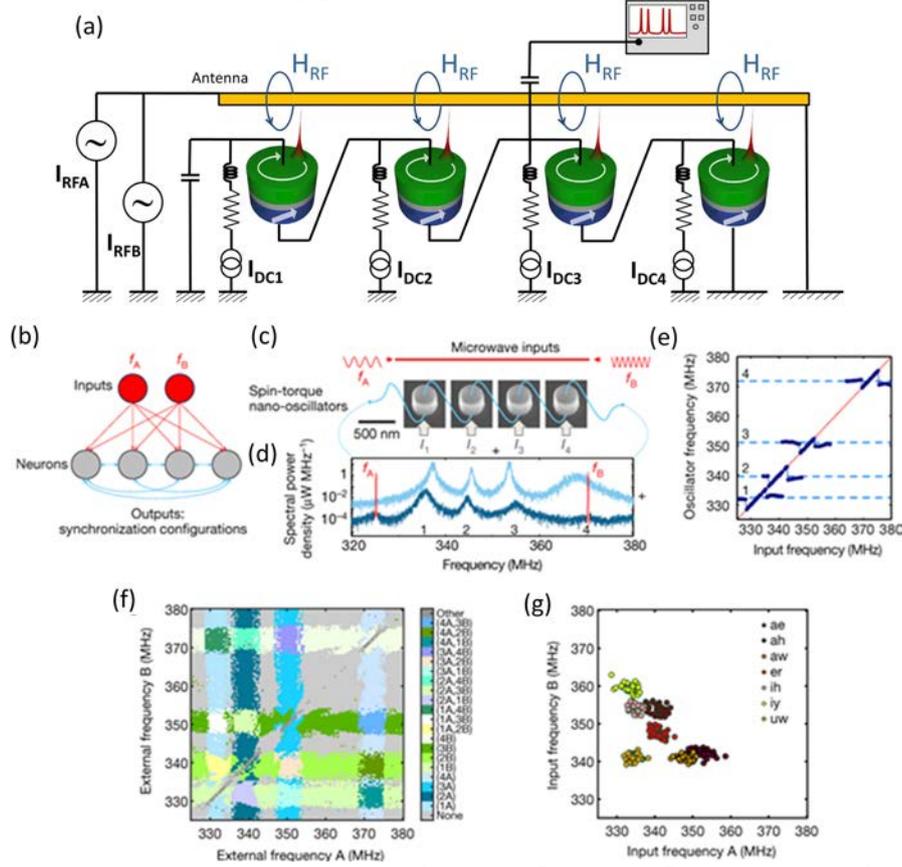

*Figure 16. The four coupled vortex nano-oscillators. $I_{RFA}$ and $I_{RFB}$ are the microwave currents injected in the strip line by the two microwave sources. $H_{RF}$ is the resulting microwave field. $I_{DC1-4}$ are the applied direct currents (a). Scheme of the emulated neural network (b). Scheme of the experimental set-up with four spin-torque nano-oscillators electrically connected in series and coupled through their own emitted microwave currents. Two microwave signals encoding information in their frequencies $f_A$ and $f_B$ are applied as inputs to the system through a strip line, which translates into two microwave fields. The total microwave output of the oscillator network is recorded with a spectrum analyzer (c). Microwave output emitted by the network of four oscillators without (light blue) and with (dark blue) the two microwave signals applied to the system. The two curves have been shifted vertically for clarity. The four peaks in the light blue curve correspond to the emissions of the four oscillators. The two narrow red peaks in the dark blue curve correspond to the external microwave signals with frequencies $f_A$ and $f_B$ (d). Evolution of the four oscillator frequencies when the frequency of external source A is swept. One after the other, the oscillators phase-lock to the external input when the frequency of the source approaches their natural frequency. In the locking range, the oscillator frequency is equal to the input frequency (e). Experimental synchronization map as a function of the frequencies of the external signals $f_A$ and $f_B$. Each color corresponds to a different synchronization state (f). Inputs applied to the system, represented in the ($f_A$, $f_B$) plane. Each color corresponds to a different spoken vowel, and each data point corresponds to a different speaker (g). Reproduced from Ref. [135] with permission of Springer Nature.*



This is an impressive result taking into account a small number of control parameters, energy efficiency and a small footprint of the device: each of the spin oscillators has a diameter of ca. 375 nm. These results demonstrate, how oscillatory dynamics can improve pattern recognition, especially in the case of dynamic patterns. These types of behaviour have been previously observed in other coupled oscillatory systems operating at the edge of chaos.[136, 137]

## 6. Transformation of signals – towards memristive cryptography

Signal and information processing involves significant security issues. From Ancient times information was considered as a high value, therefore a lot of efforts were paid to protect sensitive information. The big branch of information science is devoted to information security issues. Cryptography (from Greek *κρυπτός* "hidden, secret" and *γράφειν* "to write") is the science and technology of secure communication in the presence of adversaries trying to prevent the users from achieving privacy, integrity, and availability of data. More generally, cryptography is about constructing and analyzing protocols that prevent third parties or the public from reading private messages. Various aspects in information security such as data confidentiality, data integrity, and authentication, are central to modern cryptography.[138]

This is especially important in the era of the Internet of Things, a ubiquitous network of transmitting, computing, storage, and information retrieval devices. Therefore, in order to provide privacy and security to the users, various cryptographic techniques are used. In currently used devices most of the solutions are software-based, but the tremendous progress in unconventional and neuromimetic *in-materio* computing creates unique cryptographic solutions. It should be noted, however, that there are no complete *in-materio* cryptographic systems up to date, but some cryptographic primitives have been already implemented or at least demonstrated in numerical models. They include random number generation, identity verification via physical unclonable functions, hashing functions and ciphering/deciphering of messages.[139]

### *6.1. Chaos and random number generation*

The generation of random numbers is essential for many areas of economic importance. This type of numbers is a fundamental building block of many areas of science and technology – cryptography, stochastic modeling and probabilistic computation (Monte Carlo) as well as e-commerce, gambling, and finance-related areas.[140, 141] These areas rely heavily on random numbers and could not function without them. For this reason, it is important to develop new technologies that allow the generation of such numbers.

Generally, random numbers used in various applications are pseudo-random. They are generated through software programs based on some mathematical formula that, if run again with the same "seed", would give the same determined output. It's clear that this is an issue, especially from the point of view of cryptography, therefore, it is important to have a reliable random number generator that will produce unique sequences in each run. Compared to software solutions, a hardware approach poses a more secure option that can act as a cryptographic key. Hardware systems for generation of random numbers are referred to as True Random Number Generators (TRNG).[142] They are based on various physical processes, such as thermal noise,[143] photoelectric effect[144] or quantum phenomena[145] to name a few. Due to their fluctuating nature, these processes enable obtaining random numbers by digitizing generated signals. For example, thermal noise based TRNG operate through amplification of noise coming from resistors, avalanche diode or Zener diode. Another classic example is based on the nuclear decay of radioactive atoms, which is measured by a Geiger counter attached to a PC.[146] More advanced solutions employ chaotic laser systems mounted on the optical table.[147]

Novel TRNG technologies include systems based on memristive junctions or electronic circuits coupled with memristive elements. Employment of memristive structures is beneficial from the point of view of possible integration with existing CMOS architecture as well as their fast and energy-efficient operation. Essentially, TRNG based on memristive devices employs observed intrinsic stochasticity of resistive switching related phenomena which act as an entropy source. In



that regard, three main approaches are studied, namely – stochastic noise, stochastic switching time and stochastic switching voltage. [148] One of the first TRNG hardware implementations based on resistive switching was presented by Huang. et al.[149] Random telegraph noise (RTN) observed in low conductivity state of W/TiN/TiON/SiO$_2$/Si memristive device was used as the source of randomness. RTN is based on the random physical process of trapping/detrapping of charge carriers between e.g. defect sites in the crystal lattice. However, the received 0/1 bit distribution strongly depended on the applied potential so additional circuit elements and post-processing steps (e.g., Von Neumann correction) were necessary. On the other hand, undesirable property – from the point of view of data storage – of device-to-device and cycle-to-cycle variability of operation observed in memristive devices was used by Balatti et.al. as a source of randomness.[150, 151] Due to non-volatile character of Cu/AlO$_x$ and Ti/HfO$_x$ based memristive devices, a pair of SET/RESET voltage pulses needed to be used for generation of every bit. In both approaches cited above, additional post-processing stages were necessary to realize the true randomness of the obtained bits. Despite this fact, presented solutions failed some of the National Institute of Standards and Technology (NIST) randomness tests.[152] A slightly different approach was used by Jiang et. al. to simplify the required operations hence obtainment of random bits was more straightforward.[153] In the presented research authors used the intrinsic stochasticity of the delay time in pulse-induced switching between low and high conductive states as the basis for randomness. Moreover, used memristive structure (Ag/Ag:SiO$_2$ - operation of which is based on ionic diffusion) demonstrates volatile action in the form of self-OFF switching which reduces power consumption. The presented memristive device passed all NIST randomness tests without any post-processing. Further refinement of the concept was presented by Woo et. al., [154] where stochasticity of volatile relaxation time was also used as another source of randomness (Figure 17). What's more, operation of Pt/HfO$_2$/TiN memristive structure is based on electron trapping/detrapping which is generally faster and has lower power consumption in comparison to ionic diffusion mentioned *vide supra*.[155, 156]

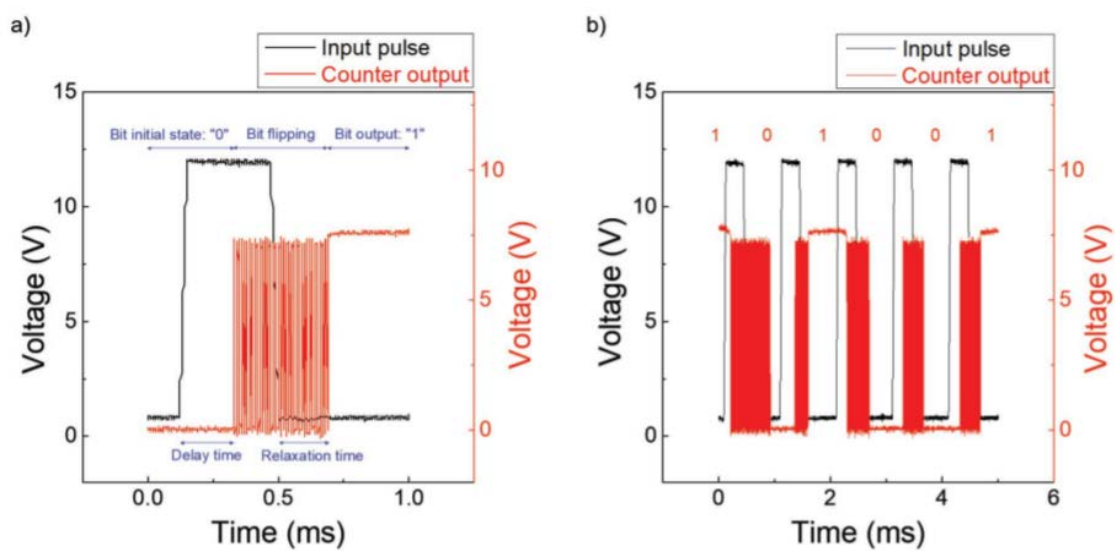

*Figure 17. Signal registered during random number generation by Pt/HfO2/TiN memristive device for a single bit (a) and a stream of bits (b). Stochasticity of delay time, relaxation time and switching of the device ("bit flipping") are shown. Reproduced from Ref. [152)] with the permission of Wiley.*

Photochromic dyes can be a platform for wetware random pattern generation. An exemplary device of this kind is based on a quartz cell (uncapped) filled with an acetone solution of a photochromic dye, e.g. 3-dihydro-1,3,3-trimethyl-8'-nitro-spiro[2H-indole-2,3'-[3H]naphth[2,1-b][1,4]oxazine][23] or 6-morpholino-3-(4-morpholinophenyl)-3-phenyl-3H-naphtho[2,1-b]pyran.[157] The cell is illuminated



at the bottom with a focused UV radiation (375 nm). Illumination induces two processes: (i) photoisomerization of colorless oxazine into colored merocyanine dye and (ii) heating of the acetone solution. The photochemical process itself is a simple photoisomerization process and does not involve any autocatalytic step, but two simple unimolecular reactions, which, in a stirred solution easily reach the photostationary state, give rise to chaotic spectrophotometric signals. The large amplitude oscillations shown in Figure 18 are observed only when enough (3 mL) photochromic solution is maintained in an uncapped cuvette and UV irradiation is carried out at the bottom of the solution. The oscillations are induced solely by the hydrodynamic instability of the solution, which is heated at the bottom and cooled form the surface via evaporation. The detailed analysis demonstrates positive values of the largest Lyapunov exponent ($\lambda = 0.045$), clearly indicating that the time series of the photochromic hydrodynamic oscillator are chaotic, and corresponding strange attractors with the fractal dimension D = 10.71.[157]

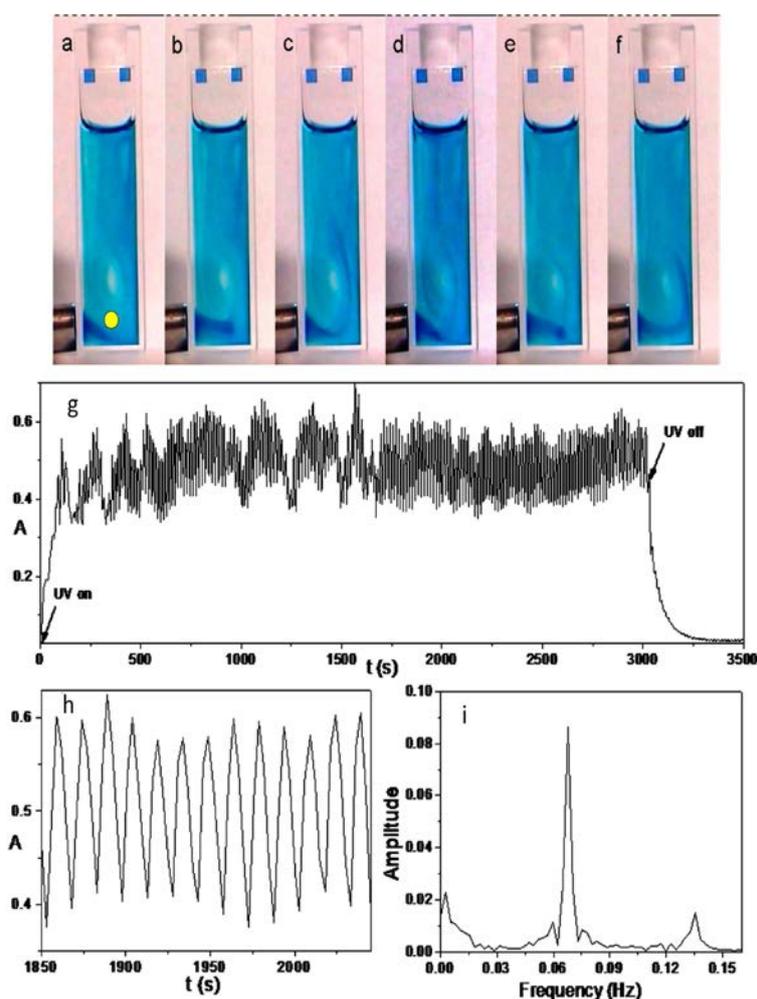

*Figure 18. Photochromic compound (concentration $9.4 \times 10^{-5}$ M) in methanol during the photocoloration stage. The snapshots in the first row are of the uncapped cuvette at 162 (a), 169 (b), 180 (c), 189 (d), 199 (e), and 209 s (f) after the beginning of UV irradiation. Oscillations recorded at 612 nm for the photochrome ($8.8 \times 10^{-5}$ M) in acetone in an uncapped cuvette. Graph (g) shows the entire dynamical evolution: the first part refers to the stage of UV irradiation (UV on), whereas the second part starts when the UV is off; (h) zoom of the kinetics showing large, regular oscillations, whose Fourier spectrum (calculated between 1850 and 2045 s) is depicted in (i). Ambient conditions: $T_{initial}$ = 298.5 K, $T_{final}$ = 298.6 K; $P_{initial}$ = 967.0 hPa, $P_{final}$ = 966.2 hPa. The height of the solution was 3.10 cm at the beginning ($h_{initial}$) and 3.00 cm at the end ($h_{final}$) of the experiment. The yellow spot in (a) indicates the place where the absorptivity of the solution was followed*



*spectrophotometrically. Reproduced from Ref.* [23] *with permission from the American Chemical Society.*

### 6.2. Physical unclonable functions

In the era of Internet of Things, one of the main challenges is to provide safe and reliable verification methods. During the recent decade, smartphones have become a universal and ubiquitous tool for a plethora of tasks, from mail and social media communication, through storage of secure information to financial operations. Since these devices often function as an authentication token for the user, the development of reliable authentication methods remains a challenging task that needs to be addressed.[158] Current authentication systems involve static random-access memory (SRAM) and non-volatile electrically erasable programmable read-only memories (EEPROMs) for the storage of the secret key. Unfortunately, these devices usually need to be combined with hardware encryption and/or digital signatures, which results in complex and expensive design.[159] This leads to increased manufacturing costs and high power consumption, as they demand a constant power supply.

A promising, less expensive alternative for authentication and key generation are Physical Unclonable Functions (PUFs). In principle, PUFs can be regarded as a black-box challenge-response system which returns a response $r = f(c)$ upon inquiry of an input challenge $c$. For a given challenge, PUF generates a unique response. Such challenge-response pairs (CRPs) are stored in a secure database.[160] When the response matches the input challenge, the device is authenticated. In this way, the authentication key is derived from the internal physical characteristics of the device instead of being stored in the digital memory and the constant power supply is no longer necessary.

In general, to be considered practical for hardware security applications, PUFs have to meet a few requirements.[161] First of all, the response of the device upon challenge query must be reproducible. Unlike TRNGs, the responses generated by the PUF should be very similar among many identical challenge inquiries.

Secondly, the response must be impossible to predict (or even completely random) and unique. The first demand can be satisfied with the manufacturing variability or with the innate variability of physical parameters of the devices: due to uncontrollable parameter variance during the production, the probability of producing two devices with identical authentication fingerprint is very low. Thereby, two different, but identically manufactured PUFs, while being inquired by the identical challenge will produce distinguishable responses, thus satisfying the requirement for the uniqueness.

Moreover, even with the complete knowledge of the PUF architecture, it should be impossible to reproduce the device. The "unclonability" of the function relies on the concept that it is impractical – both in the terms of costs and time – to reproduce the response of the device. Since the response function originates from the physical characteristics of the device, the out-of-control manufacturing variabilities guarantee that it is impossible to recreate the characteristics of the device even if the adversary could gain access to the device and get to know the space of CRPs.

Finally, the $f(c)$ and $r$ should not be bound with any trivial mathematical relationship. This prevents the adversary from predicting the functionality of the PUFs based on the known response.

PUF devices have already been implemented with the use of non-volatile memories (NVMs) and memristors. In the latter devices, the information is written by changing the resistance of the memory from one state with given conductance to another. One of the mechanisms responsible for the resistive switching phenomena relies on the creation and rupture of conducting filaments (CFs). The large concentration of defects, for example, oxygen vacancies or metallic ions injected from the active electrode, can migrate in the electric field.[162] When a positive voltage is applied to the top electrode, where the density of the defects is higher, the defects electromigrate towards the bottom electrode and change the conductivity of the device from the high-resistance state (HRS) to the low-resistance state (LRS). When a negative voltage is applied to the electrode of the device in LRS state, the field-driven migration towards the top electrode results in resistive switching to the original HRS state.



The generation and rupture of CFs that governs the transition between states are, to some extent, stochastic. Therefore, the voltage at which the switching occurs can differ between the devices and the switching cycles. It is noteworthy that this variability does not originate solely from the manufacturing process, but is inherent in the resistive switching mechanism.[163] The switching voltage and resistance variations are the main reasons restraining memristors from their application in resistive RAMs (ReRAM). However, the inability to precisely predict the current flowing through the memristors with an unknown resistive state makes them potential candidates for PUFs. One of the key requirements for PUF commercialization is a large space of the challenge-response pairs (CRPs), which cannot be obtained for simple architectures relying on single elements. The amount of CPRs can be significantly increased with the PUF implementation exploiting the cross-point arrays of memristors.

The idea for cross-point array PUFs exploits the sneak-path currents, which are a result of an indirect biasing of the array elements.[160] These currents are generally highly undesirable in crossbar memristor systems for memory applications, but due to their nature can serve as a response for an input challenge. If the resistive states in the crossbar array are randomly distributed, then the output current measured upon challenge inquiry is also random. Even if the production process introduces some spatial correlation between the elements of the array, the resistive states can be randomized by one-time programming. [160]

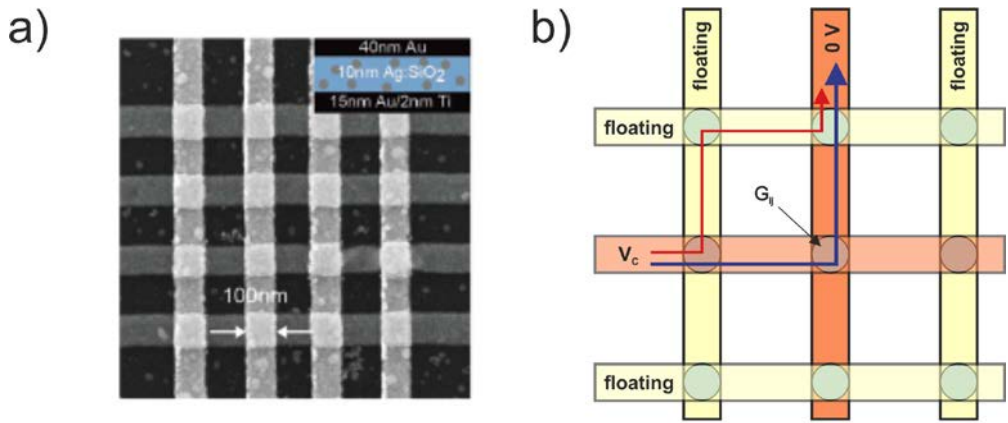

*Figure 19. PUF implementations with a crossbar array of memristors. a) SEM image of a 4x4 memristor array. b) Schematic drawing of PUF composed of a 3x3 array of memristors. Upon biasing crossbars to $V_C$ and 0 with other electrodes left floating (the challenge), the measured output current (the response) is composed of the current flowing through directly biased element $G_{ij}$ (blue arrow) and sneak-path current (red arrow). Panel a) was reproduced from Ref.[164] with permission from the Royal Society of Chemistry.*

In this implementation, the set of memristors is connected into an array with perpendicular crossbars acting as top and bottom electrodes (Figure 19a). The simplest challenge query is realized by biasing row $i$ and column $j$ with voltage $V_C$ and 0, respectively, while other electrodes are left floating.[162] The output current, besides the current flowing through element $G_{ij}$, will also include a number of sneak-path currents, being a result of an indirect biasing the array elements at the floating electrodes (see Figure 19b). Thereby, the measured current (the response) becomes a complicated function of the voltage, current flowing through $G_{ij}$ and the conductances of neighboring array elements. This concept of the challenge inquiry can be extended to an arbitrary number of electrodes to make the response function more complex.[160] In general, such PUF can be composed of any non-volatile memory elements. However, one can benefit from memristors' nonlinear I-V characteristics to improve the PUF safety.[165, 166] The unpredictable variance of physical properties in combination with the adjustable resistive states and possibility to fabricate high-dimensional stack



architectures make memristors extremely promising candidates for strong PUFs.[165]

*6.3. Hash functions*

A hash function is a cryptographic primitive, an algorithm that maps any message of arbitrary size into a fixed size (in the terms of a number of bits) number. The hash function should be a one-way function, i.e. a function which is practically infeasible to invert. The ideal cryptographic hash function has the following main properties: (i) it is deterministic, meaning that the same message always results in the same hash; (ii) it is quick to compute the hash value for any given message; (iii) it is infeasible to generate a message that yields a given hash value; (iv) it is infeasible to find two different messages with the same hash value and finally (v) a small change to a message should change the hash value so extensively that the new hash value appears uncorrelated with the old hash value.[167]

A memristor array-based hash function has been suggested by Azriel and Kvatinsky.[168] It is based on a write disturb phenomenon, a phenomenon based on parasitic currents in memristive crossbar arrays. Such arrays suffer from sneak paths – parasitic currents through unused memory cells, which distort information during the read operation and modify unselected cells during the write operation. A secure hash function is created on the basis of the state of the whole array of memristors – all memristors contribute to the hash state. Information used to generate hash is written to some cells of the array – the addresses of these cells are computed based on the message itself and a previous state of an array. Because write disturb mitigation is implemented, during the write operation, in addition to the target cell, other cells are modified. This procedure produces a unique signature of each message, being the function of a message itself and all previous messages processes in the device.

A similar hashing protocol, yet much simpler, can be implemented in dynamic photoelectrochemical devices showing short time memory features and (preferably) nonlinear light-intensity photocurrent response. In the absence of a short memory feature, the photocurrent intensity generates at a semiconducting photoelectrode can be expressed as a function of a light flux (11):[46]

$$i_{ph}(\varphi) = i_0^A \left(1 - e^{-k_A \varphi(t)}\right) - i_0^C \left(1 - e^{-k_C \varphi(t)}\right), \qquad (11)$$

where $i_0^A, i_0^C$ are saturation currents for anodic and cathodic components (which in turn strongly and nonlinearly depend on electrode potential, and $k_A$, $k_C$ are materials-dependent constants. In the simplest case saturation photocurrent intensity at applied potential $U$ can be formulated for both cathodic and anodic component as (12):

$$i_0(\varphi) = \alpha \varphi \sqrt{\frac{2e\varepsilon\varepsilon_0}{N_D}(U - V_{FB})} \qquad (12)$$

where $N_D$ is the doping density, α is the absorption coefficient and $V_{FB}$ is the flat band potential of the material. Application of additional charge trapping centers (e.g. carbon nanotubes, nanoparticles of different semiconducting material or conductive polymer coating) contributes to the fading memory feature with characteristic quenching constant ($k_q$) and corresponding memory persistence time ($t_0$). This leads to the final expression of the form (13):

$$i_{ph}(\varphi(t), t) = i_0^A \left(1 - e^{-k_A \varphi(t)}\right)\left(1 - \int_{t-t_0}^{t} e^{-k_q \varphi(t)} dt\right) - i_0^C \left(1 - e^{-k_C \varphi(t)}\right), \qquad (13)$$

where $i_0$ is given by the expression above. This makes the hashing unidirectional, due to the properties of the definite integral.



### *6.4. Memristive cryptographic systems*

In common language scrambling and encryption are sometimes used as synonyms, yet one should distinguish that the latter is usually connected with digital signal processing. Analog signals data scrambler possesses intrinsic high entropy and thus generates fully random, or at least pseudo-random, outputs. As for the actual realization of the scrambling process – the original signal is modified – usually, some of its components are transposed or inverted. For the above reasons scramblers are frequently called "randomizers". The idea behind the hardware-based implementation is to exchange numerically expensive hashing functions (vide cryptocurrencies and the whole electrical/numerical consumptions) to generator based on the intrinsic discrepancies between each individual memristive pixels/cells. Such applications have been intensively studied, yet so far theoretically.[168]

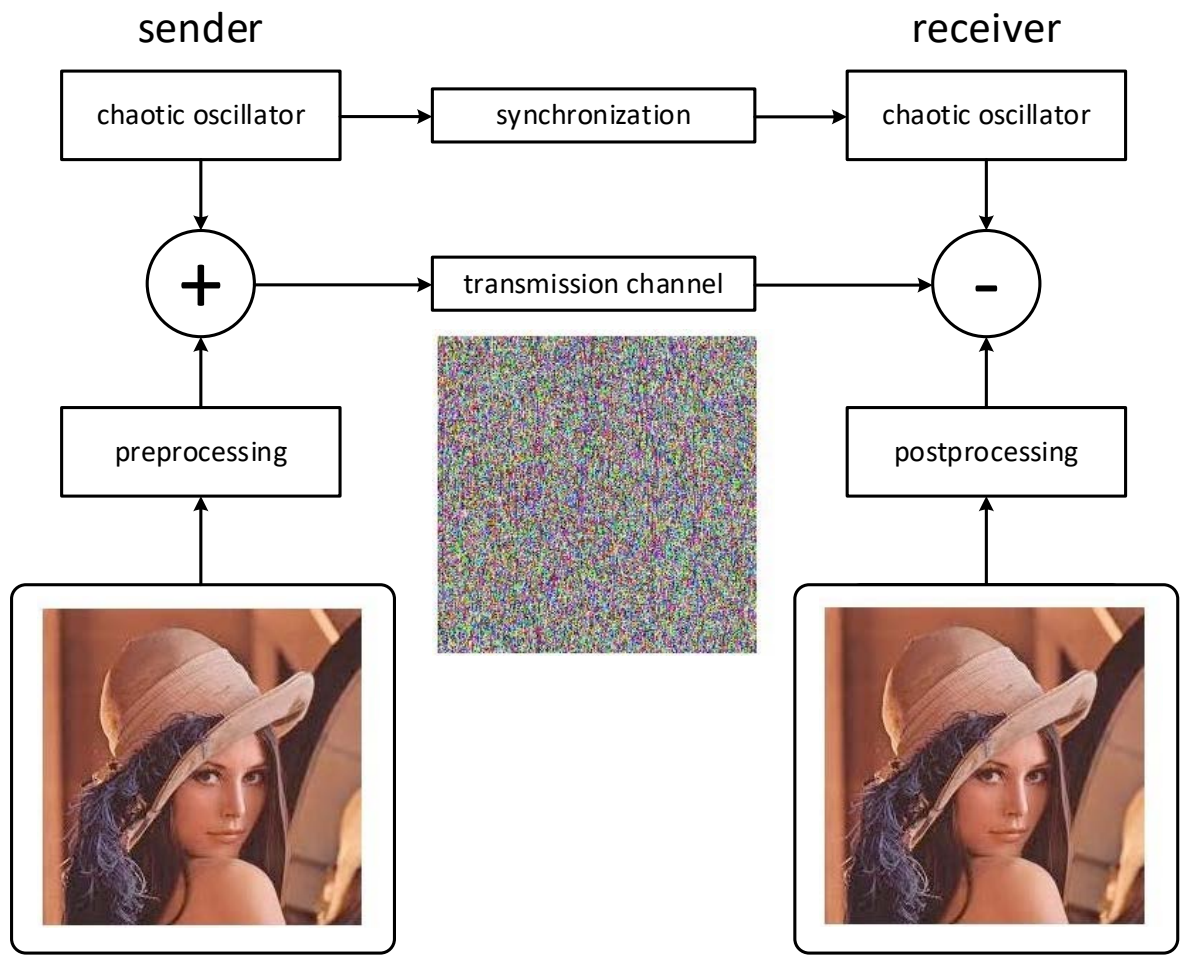

*Figure 20. A scheme of chaotic cryptography system based on two coupled Chua chaotic oscillators. Original graphical data are taken from Ref. [169] according to the CC 4.0 license.*

Data scrambling, despite being designed for analog signal, is also used in digitized information – consisting of many following 0s and 1s. Data scrambler effectively truncates the number of similar bits, making the information more lightweight. In current IT systems data scrambling functions/algorithms/concepts are also used for removing or truncating most sensitive data. Such processes are principle irreversible.

Chaotic systems can be exploited to generate apparently random sequences that serve as a basis for various cryptographic techniques.[141] One of the applications of chaotic time series is to hide an encrypted message by mixing it with a chaotic signal.[170] Chaotic dynamics are, by definition,



aperiodic, extremely sensitive to the initial conditions, and unpredictable in the long term.[171] Any chaotic signal can, therefore, mask a secret message. The sender takes the message $m(t)$ and adds it to a chaotic signal $c(t)$, creating an incomprehensible secret message $s(t)$:

$$s(t) = m(t) + c(t) \qquad (14)$$

Other operations convoluting messages with chaotic keys can be also used.

Any unauthorized adversary can detect only a chaotic signal, which sounds/looks like meaningless noise. With an appropriate, synchronized source of a chaotic signal, it is possible, however, to recover the initial message. Chaotic Chua oscillator and related memristive circuits generate truly chaotic time series,[8, 172-175] therefore they are proposed as novel cryptographic engines that can operate with analog (e.g. acoustic) and digital messages in real-time. The software implementation of this approach works spectacularly, especially on graphical data.[169]

The software model of this system was checked on a famous photo of Lena Forsén, shot by photographer Dwight Hooker, cropped from the centerfold of the November 1972 issue of *Playboy* magazine (Figure 20).[176] A set of two synchronized chaotic oscillators provide enough security in the transmission of graphical content. Application of a simple image scrambling protocol (inverting odd lines and columns followed by column permutation) and subsequent pixel-by-pixel XOR operation with the output of Chua's memristor-based oscillator yields a chaotic 'pepper-and-salt' type image. A more developed security algorithm, also based on Chua's oscillators was reported by Arpacı et al.[177] Application of hyperchaotic oscillators and additionally, a complex scrambling algorithm provides even better data protection and additionally giver better error resistance. The efficiency of this encryption process can be illustrated by a comparison of two very different images encrypted using the same protocol and the same settings of the generator (Figure 21).

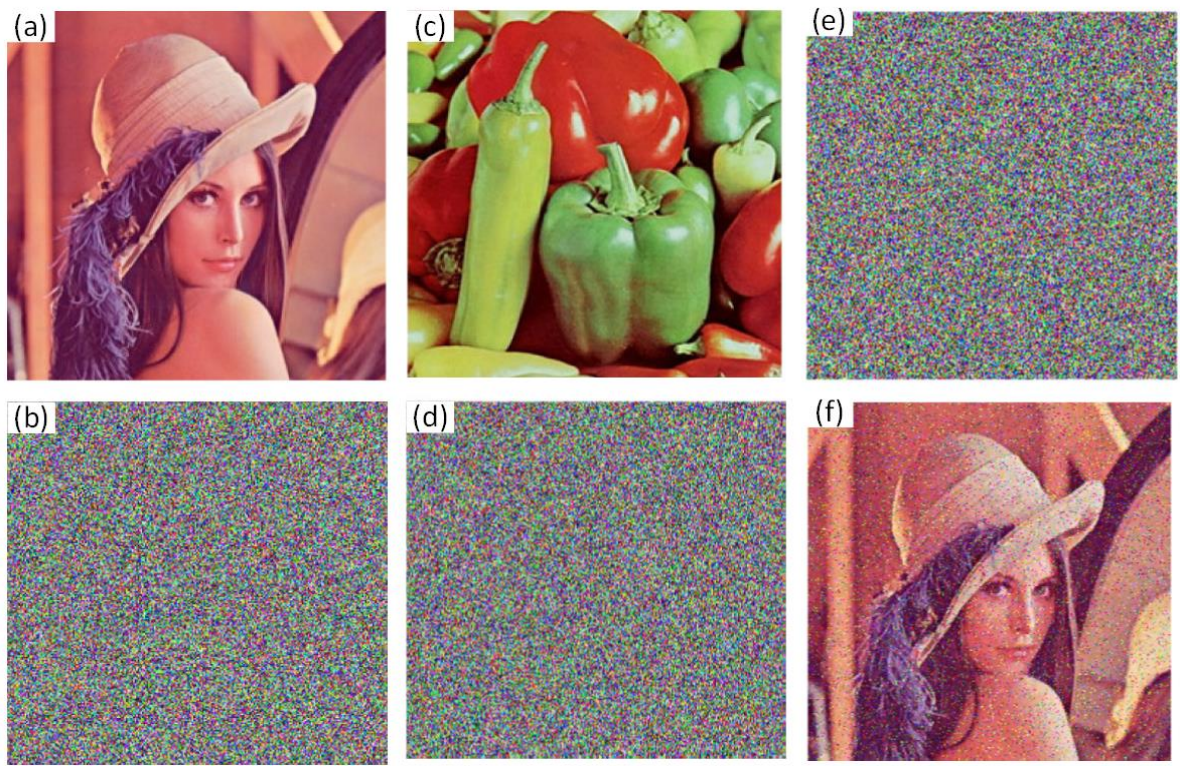

*Figure 21. Demonstration of efficiency and robustness of memristor-based chaotic cryptography: two different input data (a, c) yield very similar noise-like outputs (b, d). The addition of extra noise to the communication channel (e) still allows a successful decoding of the message (f). Reproduced from Ref. [177] with the permission of Elsevier.*

## 7. Conclusions

By drawing a simple comparison between the ability to process information *in silico* and that by biological neuronal structures, faster does not necessarily mean better. The emergence of extraordinary performance in the pattern recognition resulting from the high complexity of the nervous system and its operation at the edge of chaos[178] maximized our chances of survival and enabled the rise of the civilization as we know today. It seems to be a natural step to draw inspiration from the neural structures and transfer their functionalities into artificial systems. The imitation of human intelligence can be achieved by pursuing the idea of implementing fuzzy logic and a complex dynamic behavior with molecules and nanoscale systems and by boosting the research line of neuromorphic engineering.

The examples presented in this paper demonstrate that both wetware and memristive hardware enable efficient computation within dynamic systems with memory. This approach seems to circumvent the problem of a von Neumann bottleneck. In classical computational systems information is stored in the memory and all calculations are done within a microprocessor, therefore constant data flow slows down computation and consumes a lot of energy. Dynamic systems (both solid-state and wetware) provide a unique opportunity of computation within memory – a novel computational paradigm, which however requires new algorithms and new system architectures.

In the field of neuromorphic engineering, the exploitation of exhaustible, slow and bulky oscillatory reactions has strong limitations compared to much faster and easily scalable traditional electronic components and memristors. By implementing neuromimetic devices through unconventional chemical systems and materials, we intend to promote the development of the Chemical Artificial Intelligence (CAI).[14] The purpose of CAI is to mimic some of the performances of the human intelligence by using not software or the conventional hardware of the electronic computers based on the von Neumann architecture, but rather unconventional chemical systems in wetware or new hardware based on unconventional materials and phenomena.

As a far-fetched vision, we may predict that other oscillatory chemical reactions besides the BZ and the Orbán oscillators,[179] and after miniaturizing them by micro-beads, microcapsules, and compartmentalization through micelles, liposomes, and micro-emulsions,[180-182] the use of a wetware (i.e., solutions) rather than a hardware (solid state devices), and of electromagnetic radiation instead of chemicals or electrons, will bring two great benefits. First, in biological multicellular systems, which are complex out-of-equilibrium "soups" of chemicals, the phenomena of long-range coupling, mediated by diffusive chemicals, exist. Such long-range coupling phenomena, not possible in the solid state, enlarge the computing power of the whole system because they originate collective oscillations and waves. Finally, the encoding of information by UV-visible radiation will guarantee very fast propagation of messages, easy tunability of their content, and a wealthy code for futuristic brain-like computing machines. Finally, Chemical Artificial Intelligence will boost the development of soft robotics. Soft robots, also called "chemical robots", will be easily miniaturized and implanted in living beings.[183-185] They will interplay with cells and organelles for biomedical applications. They will become auxiliary elements of the human immune system to defeat diseases that are still incurable.

## 8. Acknowledgments

Authors thank Ewelina Wlaźlak, Andrew Adamatzky and Zoran Konkoli for numerous stimulating discussions on reservoir computing, memristive systems, unconventional computing, and signal processing. The authors acknowledge the financial support from the Polish National Science Centre within the MAESTRO project (grant agreement No. UMO-2015/18/A/ST4/00058). DP and PZ have been partly supported by the EU Project POWR.03.02.00-00-I004/16.

## 10. THE END